\documentclass{aa}
\usepackage{graphics,epsfig,txfonts,longtable}

\usepackage{natbib}
\bibpunct{(}{)}{;}{a}{}{,}

\def\rahour{\hbox{\ensuremath{^{\rm h}}}}

\newcommand{\ergcm}[1]{$\times 10^{#1}$ erg cm$^{-2}$ s$^{-1}$}
\newcommand{\oergcm}[1]{$10^{#1}$ erg cm$^{-2}$ s$^{-1}$}

\newcommand{\oergs}[1]{$10^{#1}$ erg s$^{-1}$}
\newcommand{\hcm}[1]{$\times 10^{#1}$ cm$^{-2}$}

\newcommand{\expo}[1]{$\times 10^{#1}$}

\newcommand{\OIII}{\ion{O}{iii}}
\newcommand{\SII}{\ion{S}{ii}}
\newcommand{\HI}{\ion{H}{i}}
\newcommand{\HII}{\ion{H}{ii}}
\newcommand{\Halpha}{H${\alpha}$}

\newcommand{\ltsima}{$\buildrel < \over \sim$}
\newcommand{\lsim}{\lower.5ex\hbox{\ltsima}}
\newcommand{\gtsima}{$\buildrel > \over \sim$}
\newcommand{\gsim}{\lower.5ex\hbox{\gtsima}}
\newcommand{\xmm}{{\it XMM-Newton}}
\newcommand{\chandra}{{\it Chandra}}
\newcommand{\einstein}{{\it Einstein}}
\newcommand{\gclxmm}{XMMU\,J011926.0-730134}
\newcommand{\gclxco}{011926.0-730134}
\begin{document}
 
\title{The XMM-Newton survey of the Small Magellanic Cloud
      \thanks{Based on observations with 
              XMM-Newton, an ESA Science Mission with instruments and contributions 
              directly funded by ESA Member states and the USA (NASA).}}
\author{F.~Haberl\inst{1} \and 
        R.~Sturm\inst{1} \and 
	J.~Ballet\inst{2} \and 
        D.J.~Bomans\inst{3} \and
        D.A.H.~Buckley\inst{4} \and
        M.J.~Coe\inst{5} \and
        R.~Corbet\inst{6} \and
        M.~Ehle\inst{7} \and
        M.D.~Filipovic\inst{8} \and
        M.~Gilfanov\inst{9} \and
        D.~Hatzidimitriou\inst{10} \and
        N.~La~Palombara\inst{11} \and
        S.~Mereghetti\inst{11} \and
 	W.~Pietsch\inst{1} \and
        S.~Snowden\inst{12} \and
        A.~Tiengo\inst{11,13}
       }

\titlerunning{XMM-Newton survey of the SMC}
\authorrunning{Haberl et al.}
 
\institute{Max-Planck-Institut f\"ur extraterrestrische Physik, Giessenbachstra{\ss}e, 85748 Garching, Germany\\
	   \email{fwh@mpe.mpg.de}
	   \and
	   Laboratoire AIM, CEA/DSM-CNRS-Universit{\'e} Paris Diderot, IRFU/SAp, CEA-Saclay, 91191 Gif-sur-Yvette, France
	   \and
	   Astronomical Institute, Ruhr-University Bochum, Universit{\"a}tstr. 150, 44780 Bochum, Germany
	   \and
	   South African Astronomical Observatory, PO Box 9, Observatory 7935, Cape Town, South Africa
	   \and
	   School of Physics and Astronomy, University of Southampton, SO17 1BJ, UK
	   \and
           University of Maryland, Baltimore County; X-ray Astrophysics Laboratory, Mail Code 662,
           NASA Goddard Space Flight Center, Greenbelt, MD 20771, USA
	   \and
           XMM-Newton Science Operations Centre, ESAC, ESA, PO Box 78, 28691 Villanueva de la Ca\~{n}ada, Madrid, Spain
	   \and
	   University of Western Sydney, Locked Bag 1797, Penrith South DC, NSW 1797, Australia
	   \and
	   Max-Planck-Institut f\"ur Astrophysik, Karl-Schwarzschild-Stra{\ss}e 1, 85741 Garching, Germany
	   \and
	   Physics Department, University of Crete, P.O. Box 2208, GR-710 03, Heraklion, Crete, Greece
	   \and
	   INAF, Istituto di Astrofisica Spaziale e Fisica Cosmica Milano, via E. Bassini 15, I-20133 Milano, Italy
	   \and
	   Laboratory for High Energy Astrophysics, Code 662, NASA/GSFC, Greenbelt, MD 20771, USA
	   \and
	   Istituto Universitario di Studi Superiori di Pavia, Viale Lungo Ticino Sforza 56, I-27100 Pavia, Italy
	   }
 
\date{Received 5 June 2012 / Accepted 4 July 2012}
 
\abstract{Although numerous archival \xmm\ observations existed towards the Small Magellanic Cloud (SMC) before 2009, 
          only a fraction of the whole galaxy had been covered.}
         {Between May 2009 and March 2010, we carried out an \xmm\ survey of the SMC, to ensure a complete 
         coverage of both its bar and wing. Thirty-three observations of 30 different fields with a total exposure of about 
         one Ms filled the previously missing parts.}
         {We systematically processed all available SMC data from the European Photon Imaging Camera.
	  After rejecting observations with very high background, we included 53 archival and the 33 survey observations.
	  We produced images in five different energy bands.
	  We applied astrometric boresight corrections using secure identifications of X-ray sources and  
	  combined all the images to produce a mosaic covering the main body of the SMC.}
	 {We present an overview of the \xmm\ observations, describe their analysis,
	  and summarise our first results, which will be presented in detail in follow-up papers. Here, we mainly focus on extended 
	  X-ray sources, such as supernova remnants (SNRs) and clusters of galaxies, that are seen in our X-ray images.}
	 {Our \xmm\ survey represents the deepest complete survey of the SMC in the 0.15--12.0 keV X-ray band.
	  We propose three new SNRs that have low surface brightnesses of a few \oergcm{-14} arcmin$^{-2}$ and large 
	  extents. In addition, several known remnants appear larger than previously measured at either X-rays or other wavelengths extending 
	  the size distribution of SMC SNRs to larger values.}

 
\maketitle
 
\section{Introduction}

The study of X-ray source populations and diffuse X-ray emission in nearby galaxies is of 
major importance to improve our understanding the X-ray output of more distant galaxies as well as 
learning about processes that occur on interstellar scales within our own Galaxy.
\xmm\ and \chandra\ were used to perform deep X-ray surveys of the Local Group galaxies 
M31 \citep{2005A&A...434..483P,2011A&A...534A..55S} 
and M33 \citep{2004A&A...426...11P,2006A&A...448.1247M,2011ApJS..193...31T}. 
These deep observations of M31 and M33 allow for the first time the study of 
large samples of different source classes (with limiting point-source 
luminosities of $\sim$\oergs{35}) in a galaxy other than the Milky Way. This includes,
e.g., the study of about 80 supernova remnant (SNR) candidates in M33, which is 
the largest sample of remnants detected at optical and X-ray wavelengths in any galaxy 
\citep{2010ApJS..187..495L}, 
optical novae as the major class of supersoft X-ray sources (SSSs) in M31 and M33
\citep{2005A&A...442..879P,2007A&A...465..375P,2010A&A...523A..89H,2011A&A...533A..52H} 
and the discovery of type-I X-ray bursts from neutron star X-ray binaries in M31 globular clusters 
\citep{2005A&A...430L..45P}. 

The Large and Small Magellanic Clouds (LMC and SMC), nearby neighbours of the Milky Way, have different chemical 
compositions of low metallicity, are irregular in shape, and are strongly interacting both with the Milky Way and each other.
These properties influence their star formation history and therefore, any study of stellar populations 
in the Magellanic Clouds (MCs) is particularly rewarding 
\citep[see e.g. ][ with respect to the Be/X-ray binary population of the SMC]{2010ApJ...716L.140A}. 
Their proximity makes them ideal targets for 
X-ray studies. Limiting point-source luminosities of a few 
\oergs{33} (a factor of $\sim$50 lower than for M31 and M33) are reached with \xmm\ and
extended objects like supernova remnants can easily be resolved 
\citep[at the SMC distance of 60 kpc, ][ the angular resolution of $\sim$10\arcsec\ provided by \xmm\ corresponds
to a linear size of 3 pc]{2005MNRAS.357..304H}.

Previous X-ray surveys of the MCs performed with the imaging instruments
of the \einstein\ \citep{1981ApJ...248..925L,1991ApJ...374..475W,1992ApJS...78..391W}, 
ASCA \citep[only SMC; ][]{2003PASJ...55..161Y} 
and ROSAT \citep{1999A&AS..136...81K,2000A&AS..142...41H,2000A&AS..147...75S,1999A&AS..139..277H,2000A&AS..143..391S}
satellites revealed discrete X-ray sources and large-scale diffuse emission. 
In particular, the high sensitivity and the large field of view of the ROSAT
PSPC provided the most comprehensive catalogues of discrete X-ray sources ever compiled 
and revealed the existence of a hot thin plasma in the interstellar medium (ISM)
of the MCs with temperatures between 10$^6$ and 10$^7$ K 
\citep{2002A&A...392..103S}. 

However, owing to their relatively small distance, the MCs extend over large areas on the sky and high 
spatial resolution X-ray surveys with modern instrumentation need to complete a large number of raster observations.
\chandra\ and \xmm\ observed various targets in the MCs and only for the SMC do these 
`serendipitous' surveys cover a significant part of the galaxy. The disconnected fields of the
\chandra\ Wing Survey \citep{2008MNRAS.383..330M} are distributed around the eastern wing. 
The observations available in the \xmm\ archive at the beginning of 2009 mainly covered parts of the SMC bar and 
wing with very different exposure times. 
To fill the gaps between the archival observations, we successfully applied for \xmm\ observations of 
thirty fields in the SMC, which were performed between May 2009 and March 2010. 
Together with the archival data, the new observations cover the full extent of the SMC with the 
European Photon Imaging Camera \citep[EPIC, ][]{2001A&A...365L..18S,2001A&A...365L..27T} on board \xmm.

First studies using the new \xmm\ survey data mainly focussed on individual objects belonging to the SMC. 
Two known SSSs \citep{1996LNP...472Q.299G} were observed in bright state during our \xmm\ survey. 
SMP SMC 22 \citep{1978PASP...90..621S} 
is the most X-ray luminous central star of a planetary nebula in the SMC \citep{2010A&A...519A..42M} and 
the symbiotic binary SMC\,3 is a highly variable X-ray source that was seen at its highest intensity state observed so far
during our survey \citep{2011A&A...529A.152S}. 
A faint SSS identified with a Be star could be the first Be/white dwarf system discovered in the SMC \citep{2012A&A...537A..76S}.
New Be/X-ray binaries were discovered in outburst during the \xmm\ survey observations. For two of them, 
X-ray pulsations were discovered in the EPIC data \citep{2011A&A...527A.131S,2011MNRAS.414.3281C}, while two new transients
show all the characteristics of Be/X-ray binaries, but no significant pulsations were detected \citep{2012MNRAS.424..282C}. 
Another new Be/X-ray binary was discovered in a search for highly absorbed X-ray binaries in the EPIC data 
\citep{2011A&A...532A.153N}.
\citet{2011A&A...530A.132O} combined all available \xmm\ data of the SNR IKT\,16 to study its morphology and X-ray spectrum. 

Here, we present the first EPIC mosaic images from the survey and focus on extended X-ray sources such as SNRs in the SMC and 
galaxy clusters in the background. A detailed analysis of the images concerning the detection of point sources is presented in 
\defcitealias{Sturm_cat}{SHP12}\citet[][ hereafter \citetalias{Sturm_cat}]{Sturm_cat}, which resulted in 5236 detections of 3053
individual sources.

\section{Observations and data analysis}

\begin{figure*}
 \sidecaption
   \resizebox{12cm}{!}{\includegraphics[clip=]{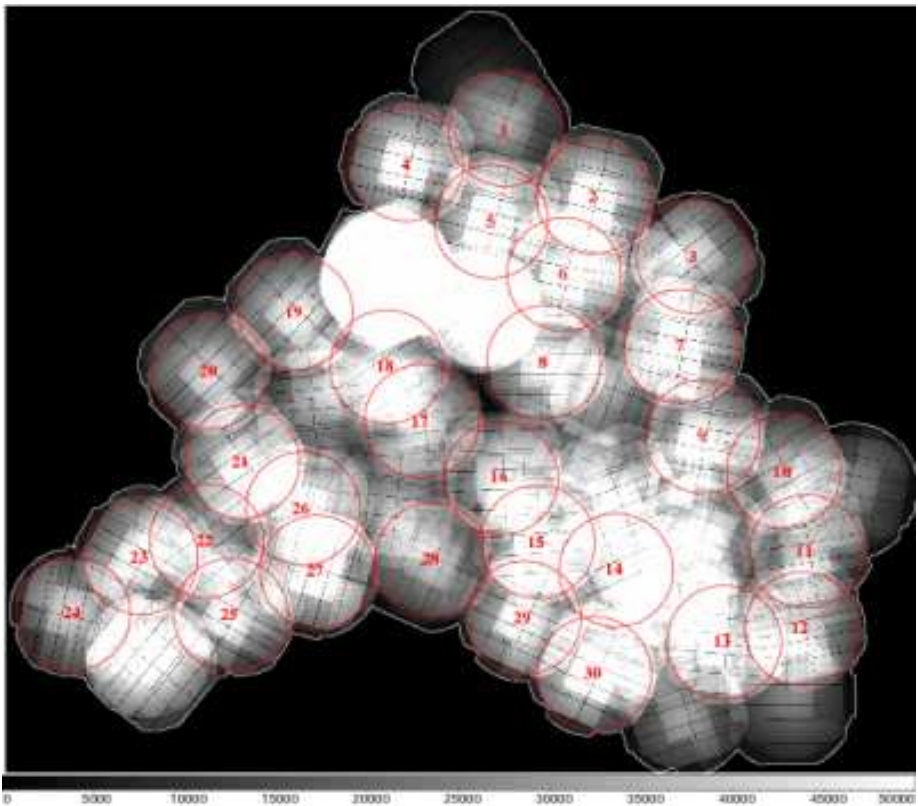}}
   \caption{Combined EPIC exposure map of the SMC area. The pn and MOS exposure maps account for vignetting  
   and for illustration are summed up with MOS1 and MOS2 each weighted with a factor of 0.4 to account for their relative sensitivity.
   The 30 numbered circles (radius 13') mark the survey fields. 
   }
   \label{fig-expo}
 \end{figure*}

The observations of the 30 SMC-survey fields were performed with all EPIC instruments in full-frame CCD readout mode with a 
73~ms frame time for pn and 2.6~s for MOS. For the pn camera the thin and for the MOS cameras the medium optical blocking 
filters were used. The locations of the fields in the SMC are indicated in Fig.~\ref{fig-expo}.
We combined the survey data with all archival imaging data available in June 2010. We selected fields with pointing 
directions within 180\arcmin\ around R.A. = 1\rahour\ and Dec. =  -73.5\degr, but excluded 
five fields located very north and south that do not overlap with the other fields. We rejected observations severely 
affected by high background (see below). This finally resulted in 53 archival observations (including the calibration target 
1E0102.2-7219), which we used for our image analysis.
A short summary of these observations is provided in Table~\ref{tab-obs}, as online material. 
We refer to Table~1 of \citetalias{Sturm_cat} for more details on the instrumental 
set-ups and net exposure times of these observations and additional observations, which were used for the production of the 
point-source catalogue. 

We used the \xmm\ Science Analysis System (SAS), version 10.0.0\footnote{http://xmm.esac.esa.int/sas/} for the analysis 
of the EPIC data. 
To remove times of high background we inferred good time intervals (GTIs) from the background light curves created by 
{\tt epchain} and {\tt emchain} using the SAS task {\tt tabgtigen} with thresholds of 8 and 2.5 cts ks$^{-1}$ arcmin$^{-2}$ 
for EPIC pn and EPIC MOS, respectively. We combined the GTIs using only common time intervals when data from all detectors 
were available, but included GTIs for a detector when others were not taking data. In this way, e.g., we ensure to include 
GTIs of MOS from the beginning of the observations before pn starts observing or observations when pn was partly in a 
non-imaging mode. 
Most of the survey observations were affected by relatively little background 
flaring activity as can be seen by the final net exposure times listed in the last column of Table~1 of \citetalias{Sturm_cat}. 
From the survey observations, only observations 0601210101 and 0601211701 received less than 60\% of their requested exposure times. Both fields were 
observed again, together with field 0601212801, which lost exposure due to a ground station problem (resulting in observations 
0601213201, 0601213301 and 0601213401). Together with the archival data, this resulted in a total number of 
83 EPIC pn and 86 EPIC MOS observations covering the bar and wing of the SMC, which we used for our analysis 
(in timing mode, pn provides no imaging data, while for MOS, the outer CCDs are still in imaging mode, reducing 
the number of pn imaging exposures). 

\begin{figure}
   \resizebox{\hsize}{!}{\includegraphics[clip=]{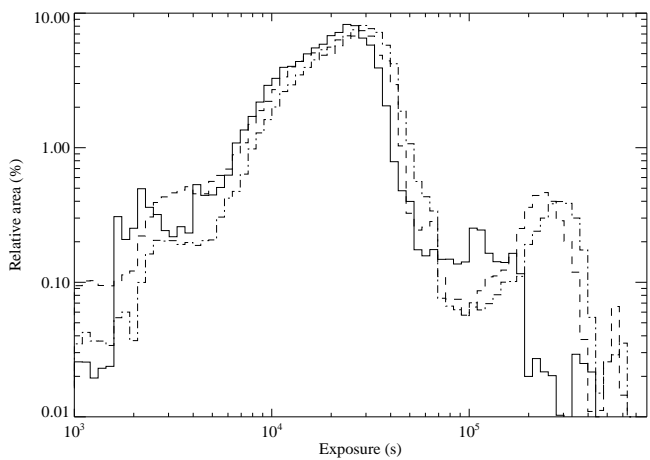}}
   \caption{Distribution of the EPIC exposures in the SMC area covered by the \xmm\ observations. 
            Full, dashed, and dash-dotted lines indicate pn, MOS1, and MOS2, respectively.
            }
   \label{fig-histexpo}
 \end{figure}

\begin{figure}
   \resizebox{\hsize}{!}{\includegraphics[clip=]{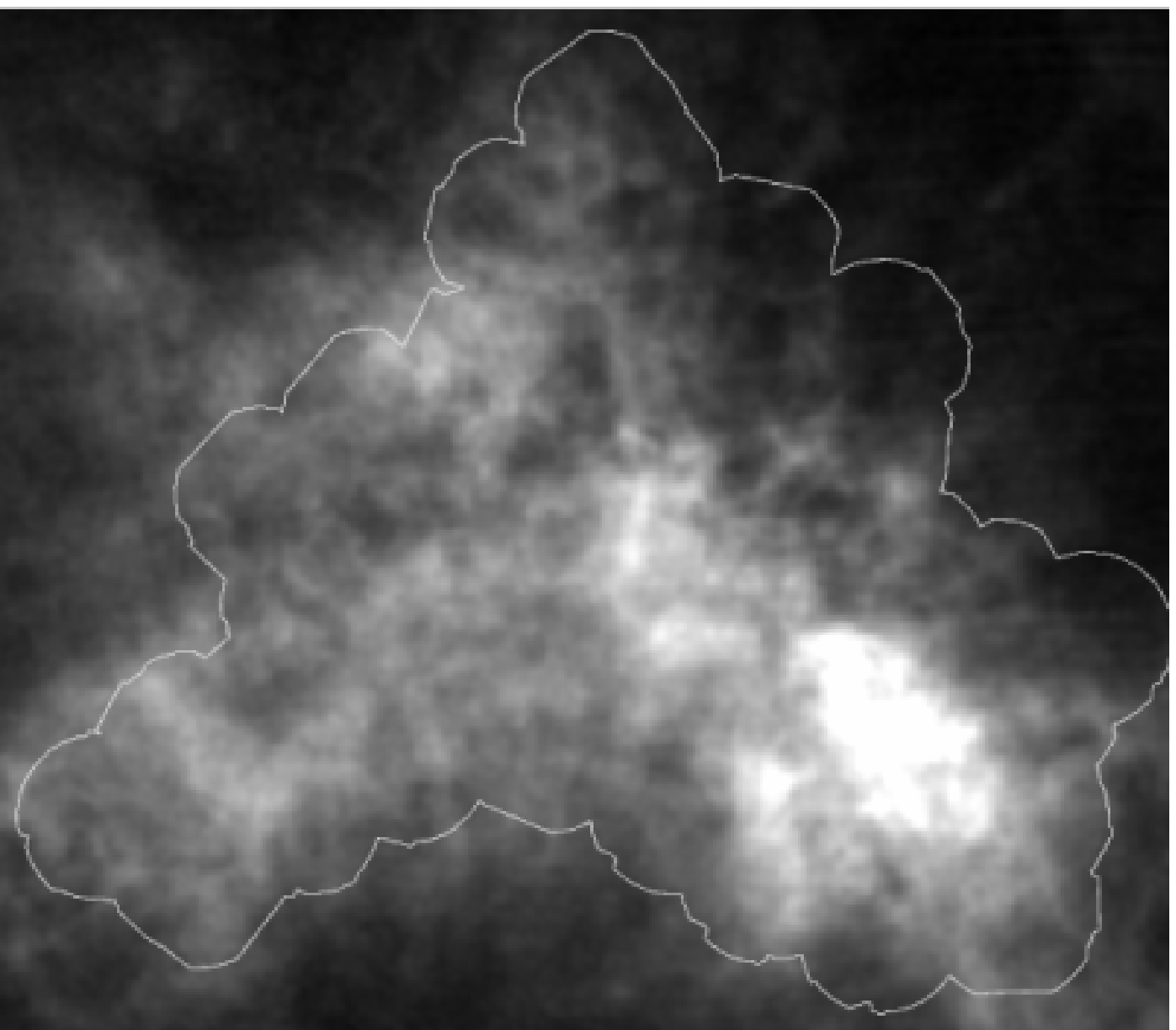}}
   \caption{Location of the \xmm\ survey with respect to the distribution of \HI\ gas in the SMC \citep[from ][]{1999MNRAS.302..417S}.
            }
   \label{fig-HImap}
 \end{figure}

\begin{figure}
   \resizebox{\hsize}{!}{\includegraphics[clip=]{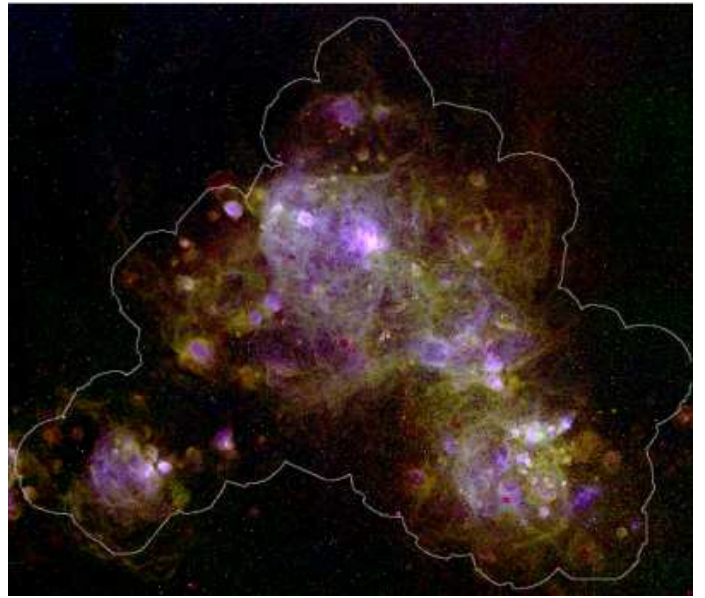}}
   \caption{Continuum-subtracted MCELS image where red, green, and blue correspond to the 
         \Halpha, [\SII], and [\OIII] emission line intensities. The contour surrounds the \xmm\ survey region. 
            }
   \label{fig-MCELS}
 \end{figure}

In a first run, we performed a source detection analysis on each observation using a maximum likelihood technique 
(SAS meta task {\tt edetect\_chain}) for 15 images from the three EPIC instruments 
in the five different energy bands 0.2-0.5 keV, 0.5-1.0 keV, 1.0-2.0 keV, 
2.0-4.5 keV and 4.5-12 keV. The resulting source lists from each observation were 
correlated with the optical catalogue from the Magellanic Clouds Photometric Survey (MCPS) of \citet{2002AJ....123..855Z}.
For astrometric boresight corrections, we selected secure identifications of optical counterparts to the X-ray sources 
(in most cases high mass X-ray binaries, HMXBs, but also foreground stars and background AGN). 
The inferred shifts in right ascension and declination were then applied to the attitude file of each observation 
and the processing repeated. In this way, astrometrically aligned images used for the 
full SMC mosaic and corrected source positions for a catalogue were produced. All images were corrected for exposure and the 
images from EPIC pn were corrected for out-of-time events to remove readout streaks. 
The MOS data were checked for noisy CCDs \citep{2008A&A...478..575K} using the SAS task {\tt emtaglenoise}. 
The data of noisy CCDs were completely rejected for the creation of the mosaic images.
A combined EPIC exposure map is presented in Fig.~\ref{fig-expo}, which covers in total 5.5 square degrees. 
In most parts of the covered area, the exposure is around 25 ks (EPIC pn only) as demonstrated in Fig.~\ref{fig-histexpo}. 
From the total covered area, 68\% (60\%, 59\%) was observed with pn (M1, M2) exposures between 10 and 30 ks. 
An extremely long exposure was reached around the supernova remnant 1E0102.2-7219 in the north-east, which is used as a 
calibration target for \xmm\ with regular observations every half a year. The maximum exposures are 427.7 ks, 585.4 ks, 
and 598.9 ks for EPIC pn, MOS1, and MOS2, respectively. For better orientation and comparison with the emission at other 
wavelengths, we mark in Fig.~\ref{fig-HImap} the surveyed area on the \HI\ map of the 
SMC obtained by  \citet{1999MNRAS.302..417S} and in Fig.~\ref{fig-MCELS} we show the corresponding area on an image obtained 
from the Magellanic Clouds Emission Line Survey\footnote{MCELS : http://www.ctio.noao.edu/mcels/}.

\begin{figure*}
   \resizebox{\hsize}{!}{\includegraphics[clip=]{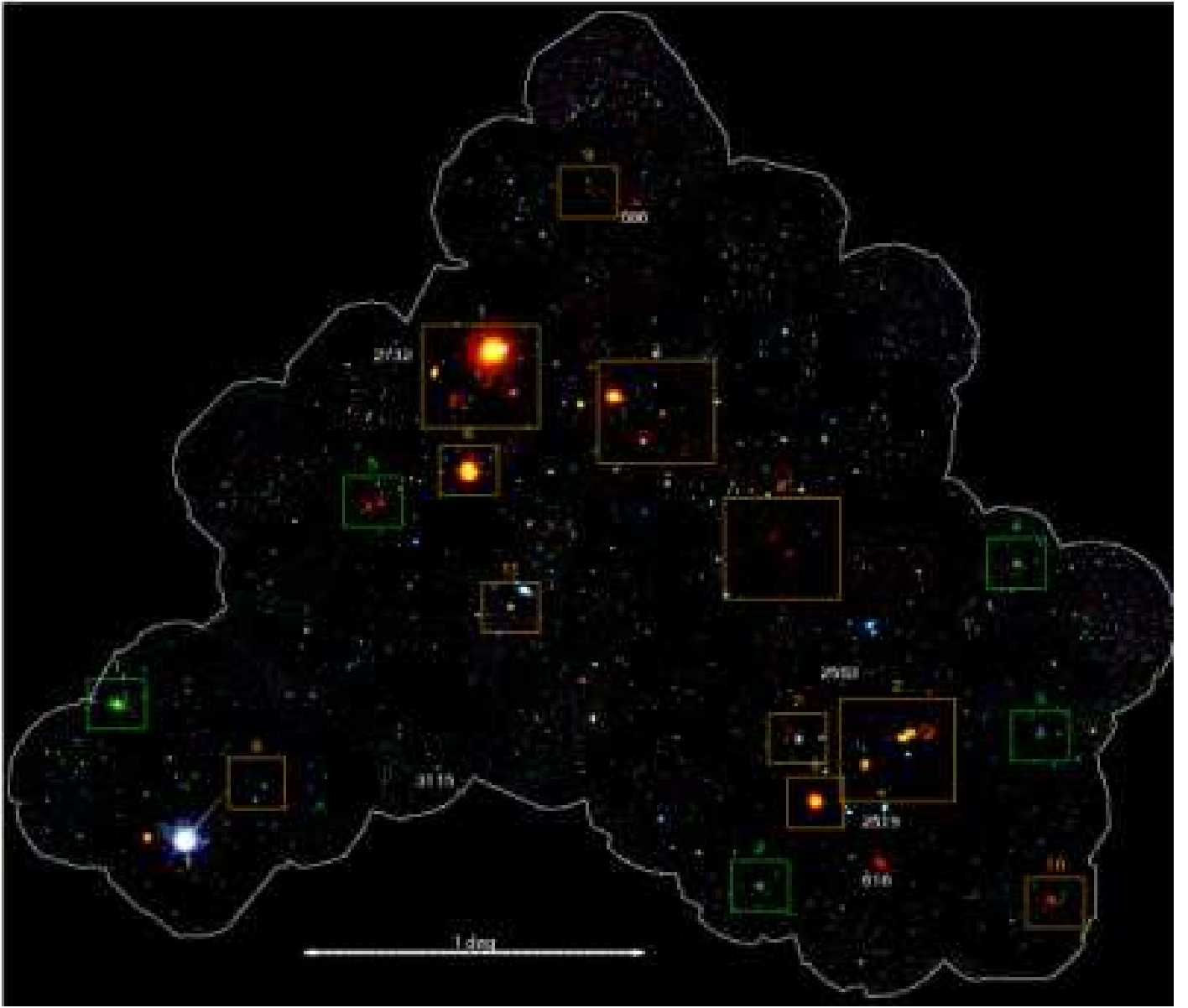}}
   \caption{Combined EPIC pn and MOS mosaic image of the SMC region centred at R.A. = 01:00:30, Dec. = -72:28:00 (J2000). The RGB colour image is composed 
   of the three energy bands 0.2$-$1.0 keV (red), 1.0$-$2.0 keV (green), and 2.0$-$4.5 keV (blue)
   (exposure-corrected and for EPIC pn out-of-time event subtracted, but not detector-background subtracted and not vignetting-corrected). 
   The boxes mark fields with SNRs (orange) and galaxy clusters (green), which are shown in more detail 
   in Figs.~\ref{fig-snrima} and \ref{fig-gclima}. Several X-ray sources, which are mentioned in the text, are marked in white with their 
   catalogue number from \citet{Sturm_cat}: 686 the planetary nebula SMP SMC 22; 616 the symbiotic binary SMC\,3; 1 the super-giant HMXB SMC\,X-1; 
   2519, 2563, 2732, and 3115 new Be/X-ray binaries.
   }
   \label{fig-rgbima}
 \end{figure*}

In Fig.~\ref{fig-rgbima}, a combined mosaic from three energy bands is presented as a colour (RGB) image.
Sources with a soft X-ray spectrum appear as red sources while hard X-ray spectra are indicated by blue colours. 
Owing to their extremely soft X-ray spectrum with counts practically all below 700 eV, SSSs
appear as very red unresolved sources in the RGB image. Two bright SSSs mentioned in the introduction,
SMP\,SMC\,22 and SMC\,3, are located in the north and south west end of the SMC bar, respectively. Owing to the 
large off-axis angle in the high state observation the latter source appears blurred. 
Both SSSs are marked in Fig.~\ref{fig-rgbima} by their catalogue numbers (686 and 616, respectively).
The HMXBs show the hardest spectra and appear blue in the image. The brightest is SMC\,X-1, which is 
south east in the eastern wing. Its location and those of the newly discovered Be/X-ray binaries from the introduction 
are also indicated in Fig.~\ref{fig-rgbima}.

To investigate the faint large-scale diffuse emission (whose analysis is ongoing and whose results will be published elsewhere) 
and SNRs in the SMC, we subtracted the detector background using data from observations with the 
filter wheel closed\footnote{available at http://xmm2.esac.esa.int/external/xmm\_sw\_cal/back\-ground/filter\_closed/index.shtml 
in their original version from 2010} 
\citep[scaled to the intensity measured in the shaded detector corners, following the work of ][]{2008A&A...489.1029B}
and corrected for the telescope vignetting. Because no data in the corners are available in EPIC pn small- and large-window mode observations, 
these had to be excluded from our analysis.

\subsection{Supernova remnants}
\label{sect-snr}

\begin{figure*}
  \resizebox{0.48\hsize}{!}{\includegraphics[clip=]{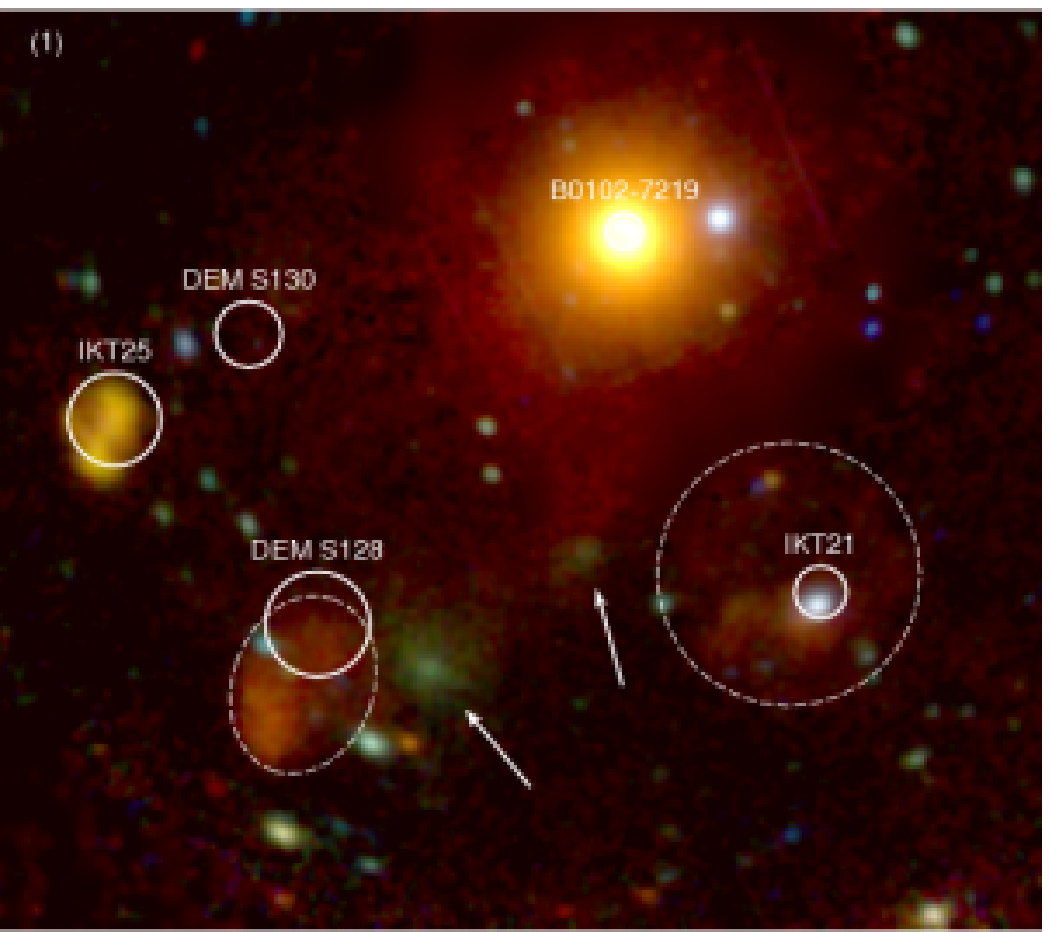}}
  \resizebox{0.48\hsize}{!}{\includegraphics[clip=]{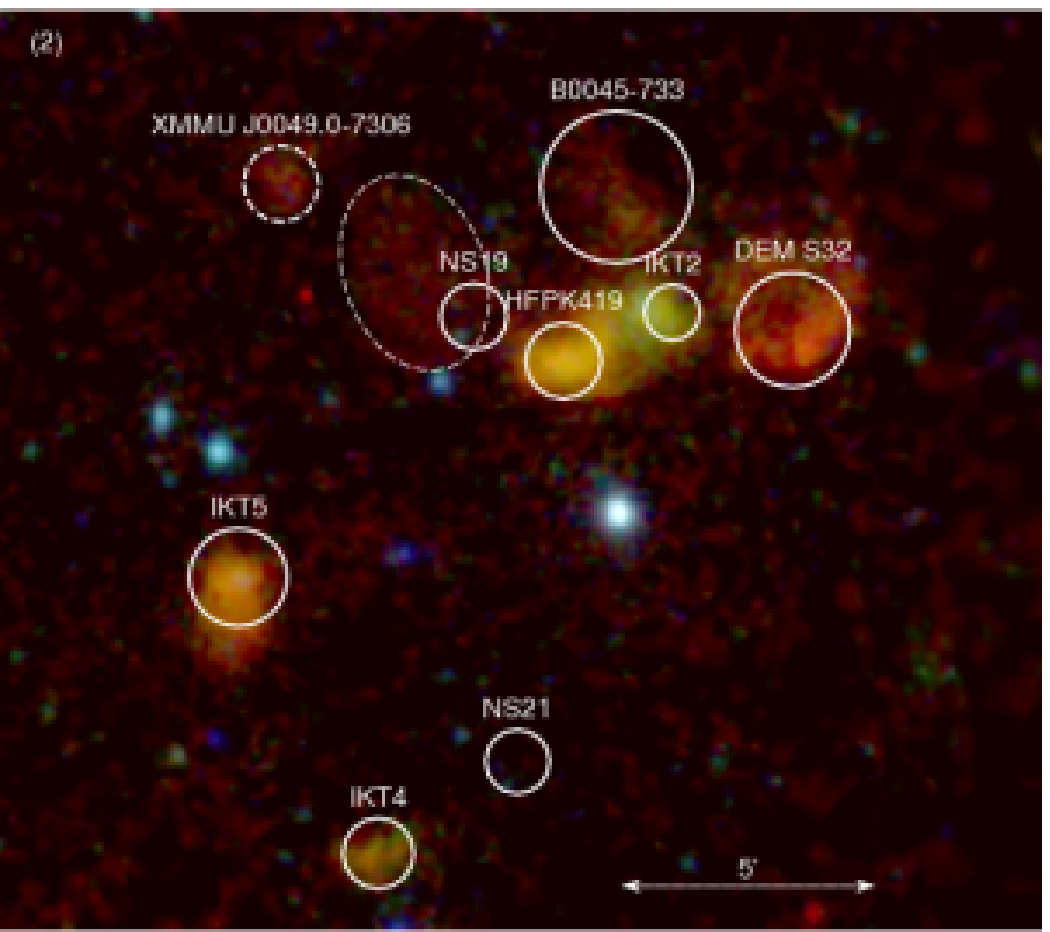}}
  \\ \noindent
  \resizebox{0.48\hsize}{!}{\includegraphics[clip=]{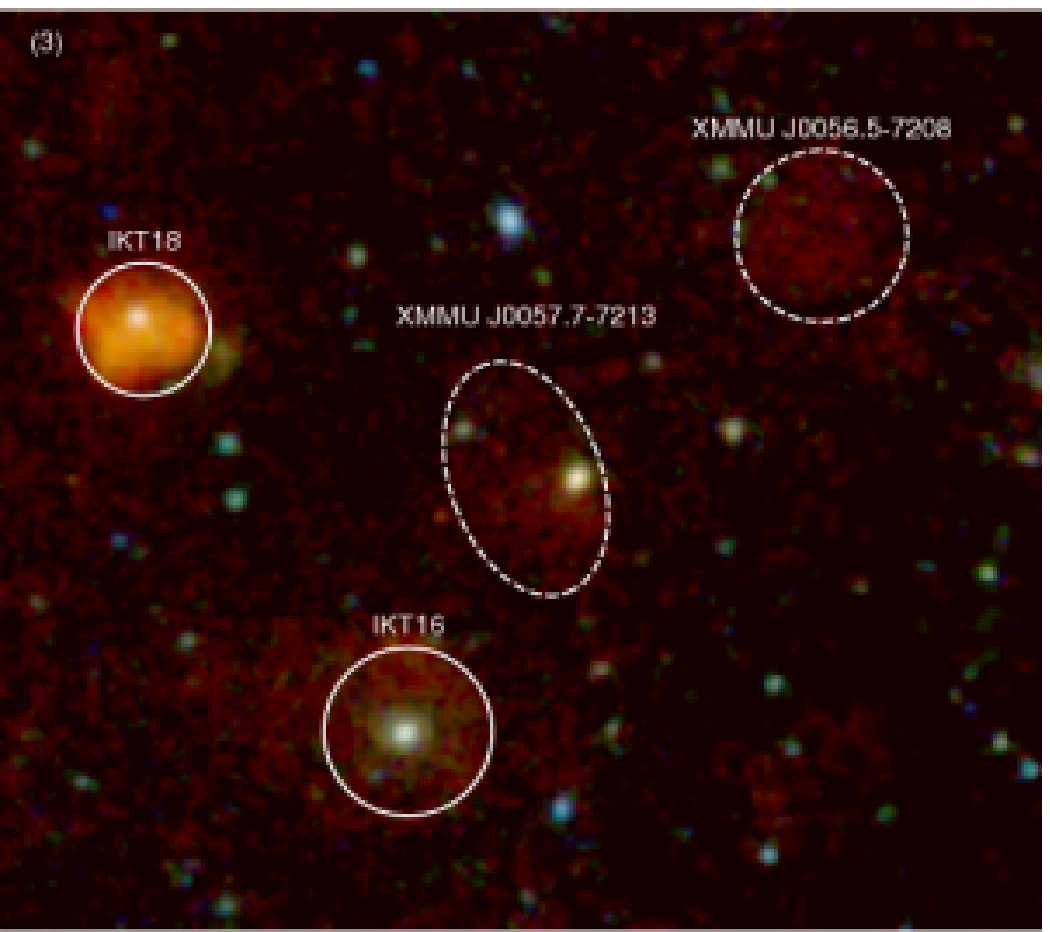}}
  \resizebox{0.48\hsize}{!}{\includegraphics[clip=]{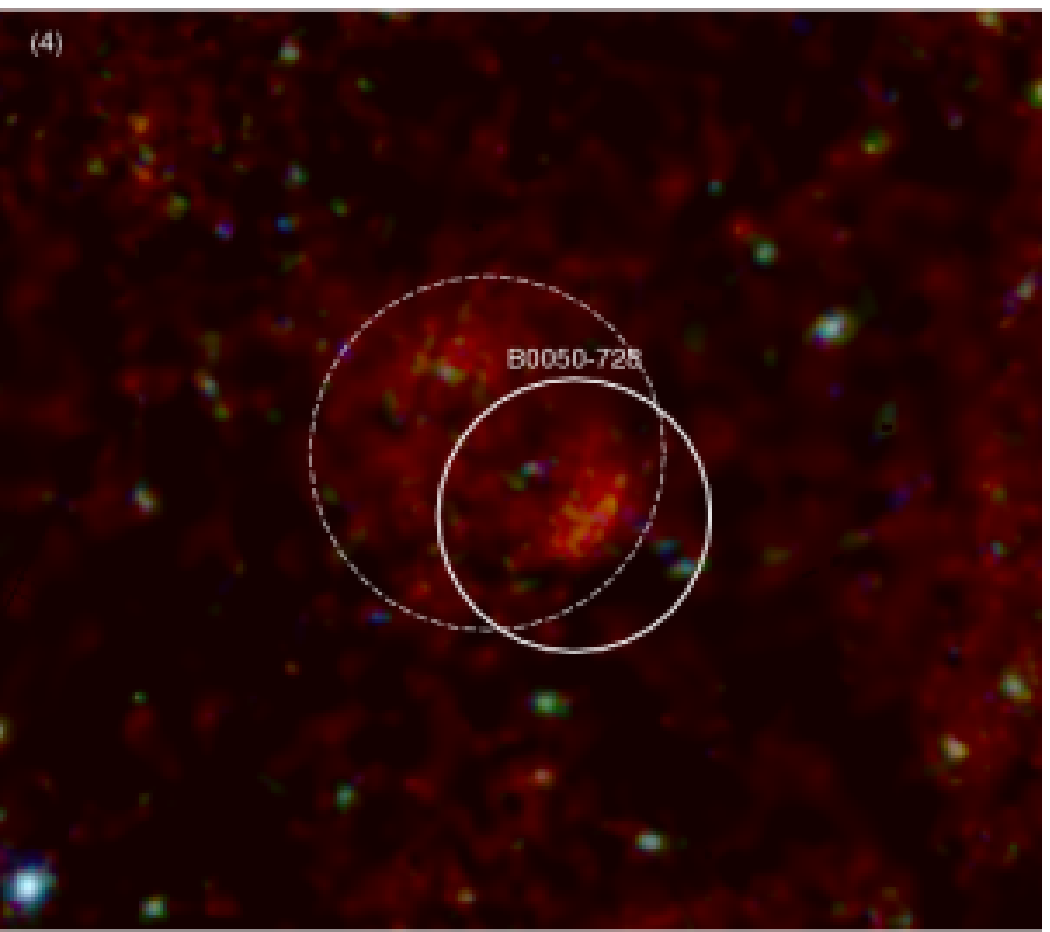}}
  \\ \noindent
  \parbox[b]{0.999\hsize}{
    \resizebox{0.238\hsize}{!}{\includegraphics[clip=]{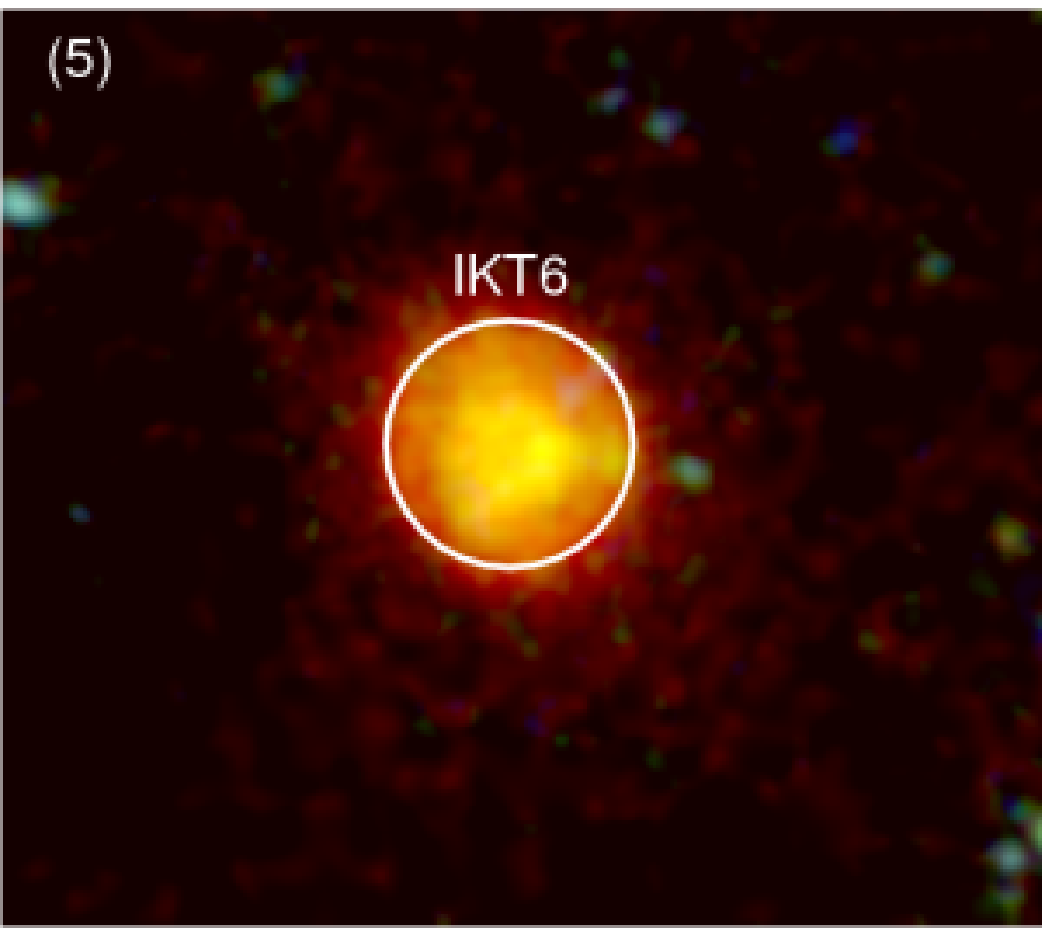}}
    \resizebox{0.238\hsize}{!}{\includegraphics[clip=]{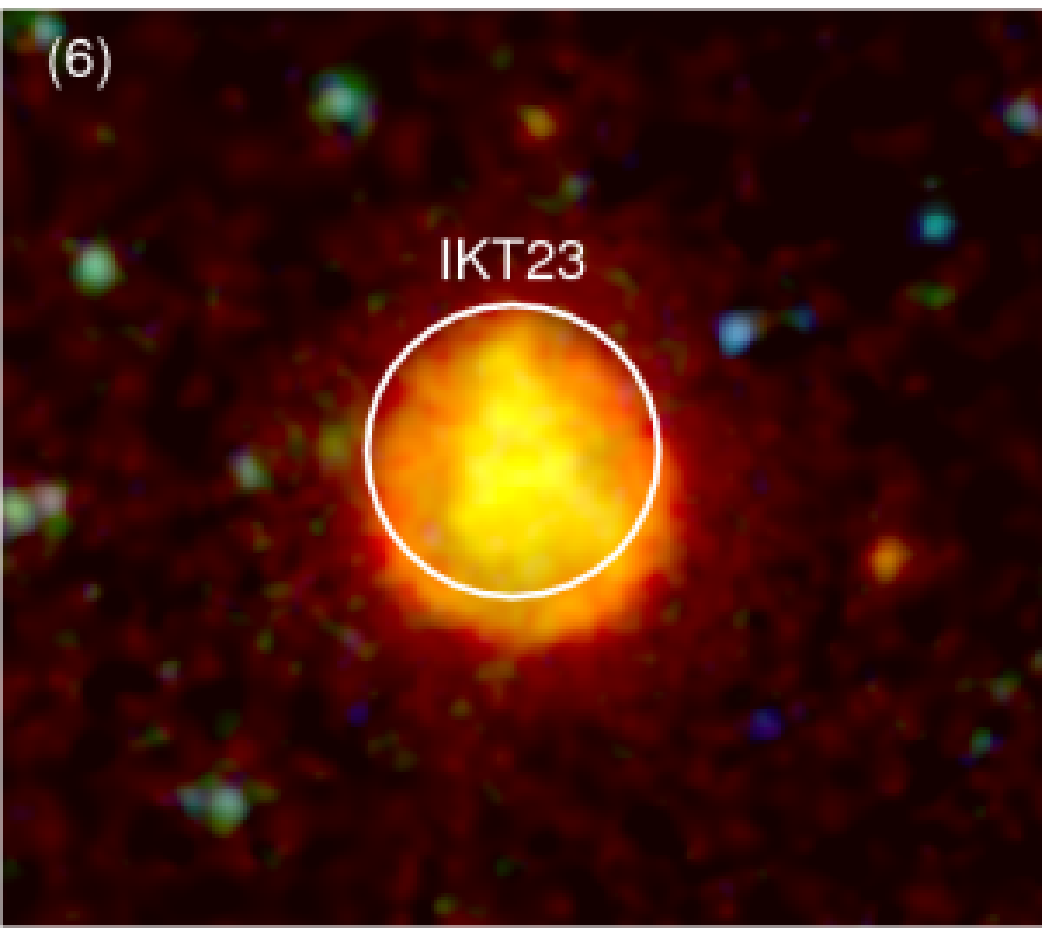}}
    \resizebox{0.238\hsize}{!}{\includegraphics[clip=]{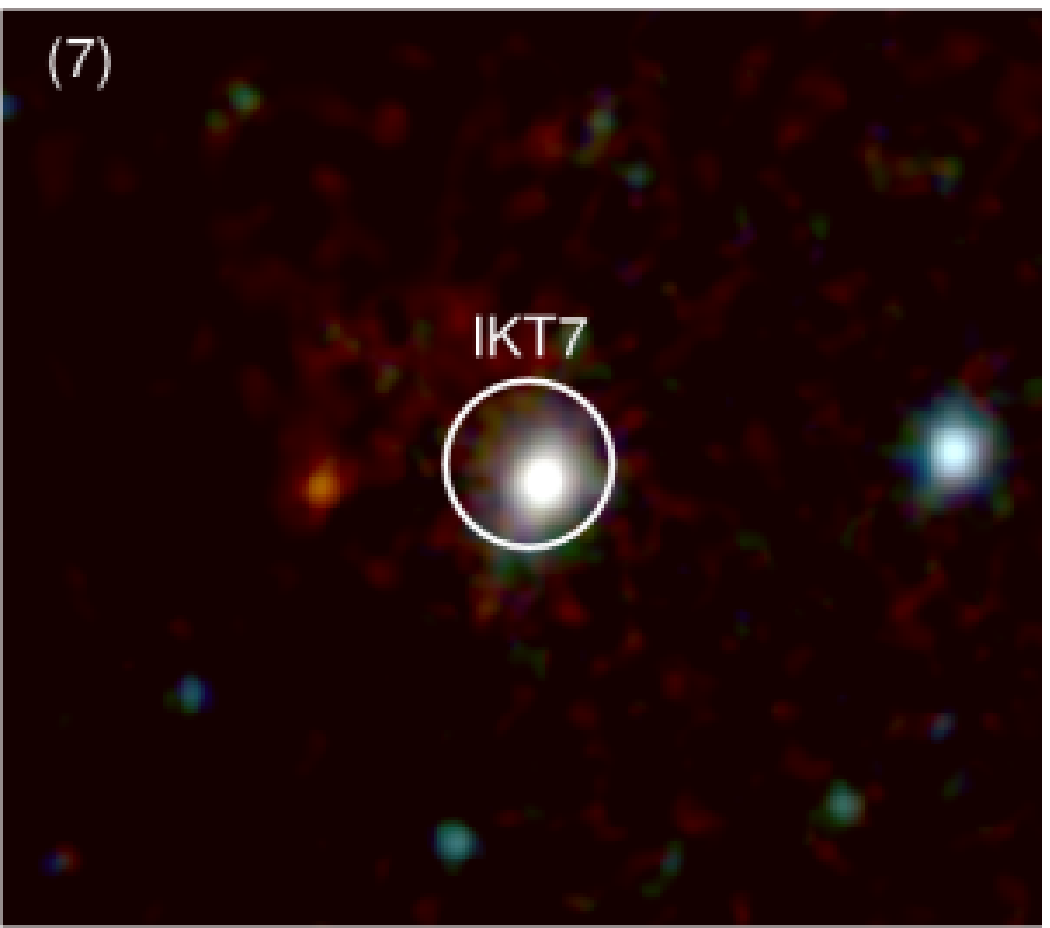}}
    \resizebox{0.238\hsize}{!}{\includegraphics[clip=]{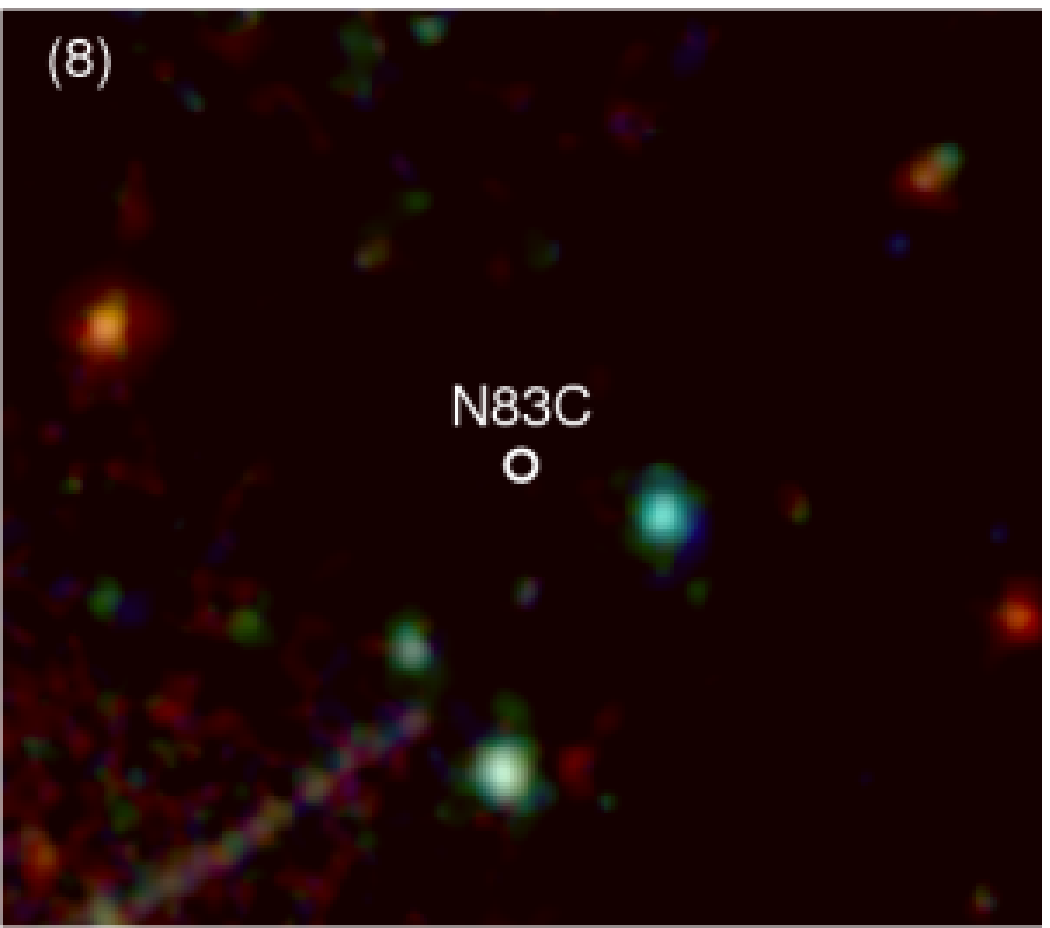}}
    \\ \noindent
    \resizebox{0.238\hsize}{!}{\includegraphics[clip=]{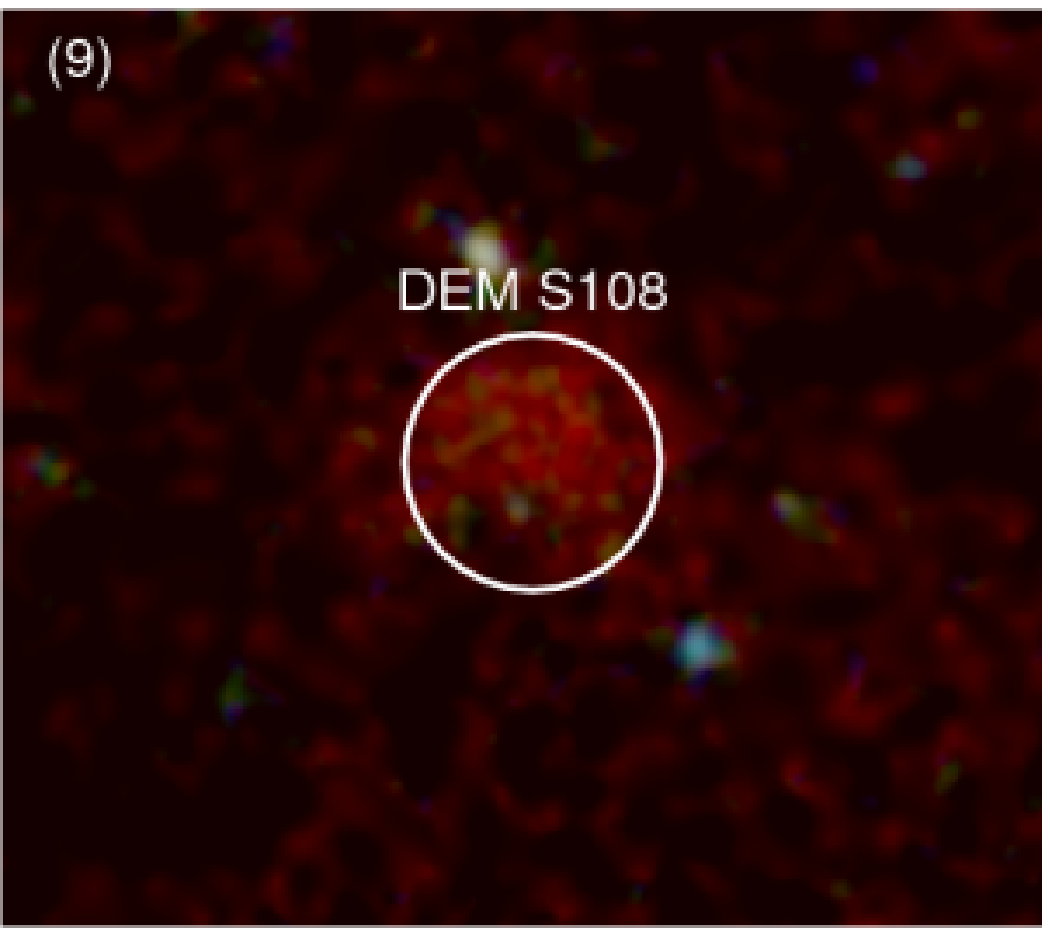}}
    \resizebox{0.238\hsize}{!}{\includegraphics[clip=]{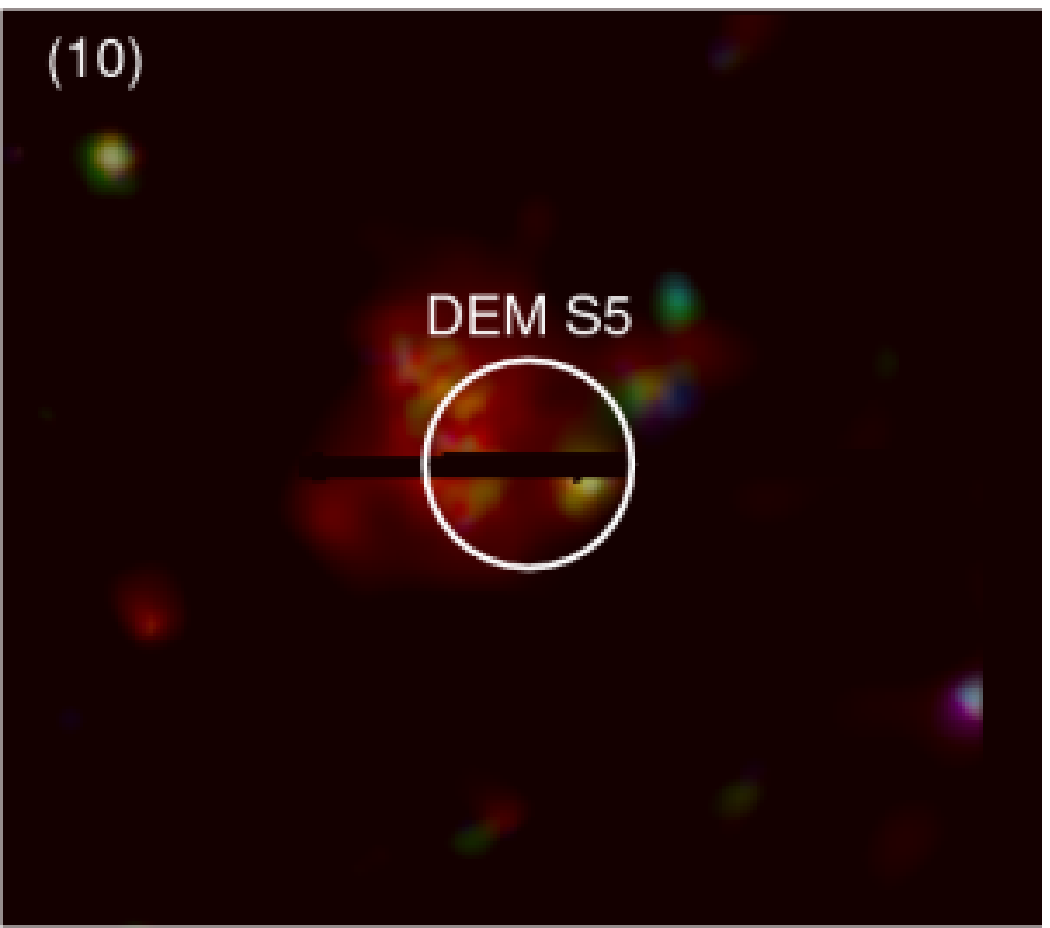}}
    \resizebox{0.238\hsize}{!}{\includegraphics[clip=]{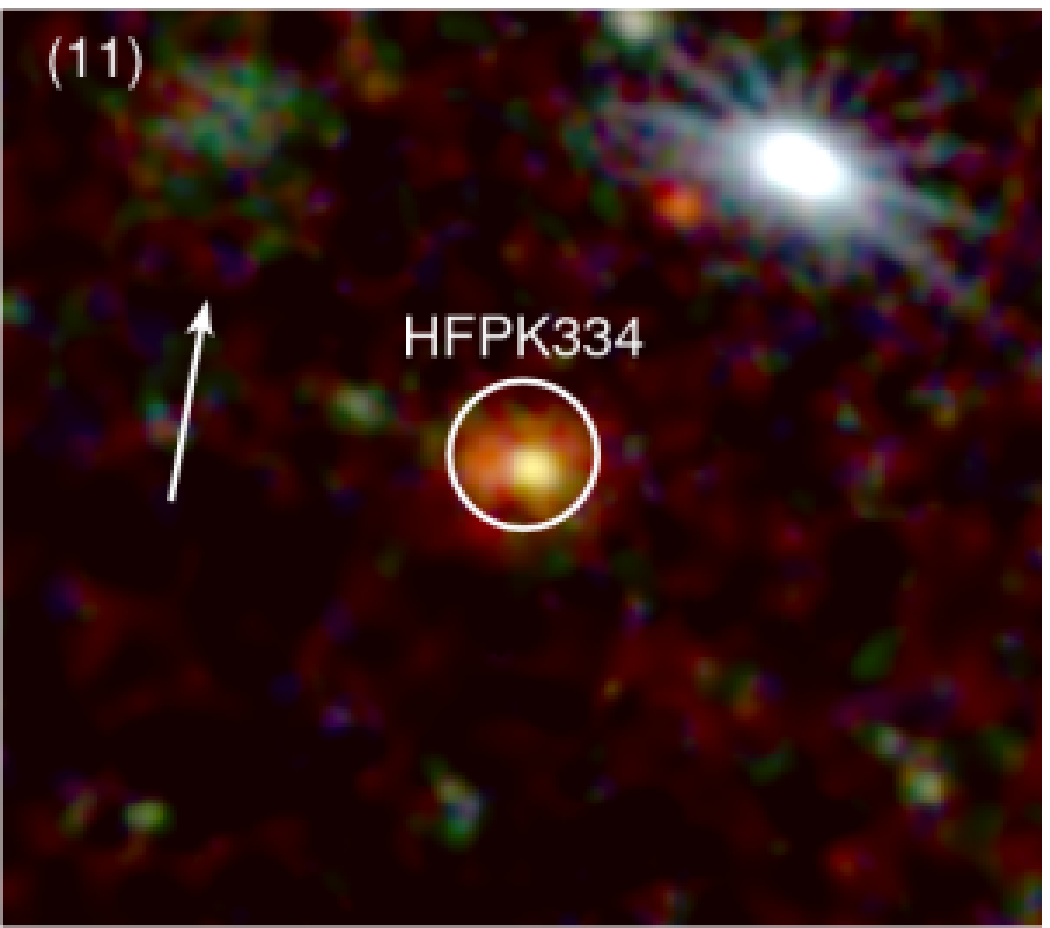}}
    \resizebox{0.238\hsize}{!}{\includegraphics[clip=]{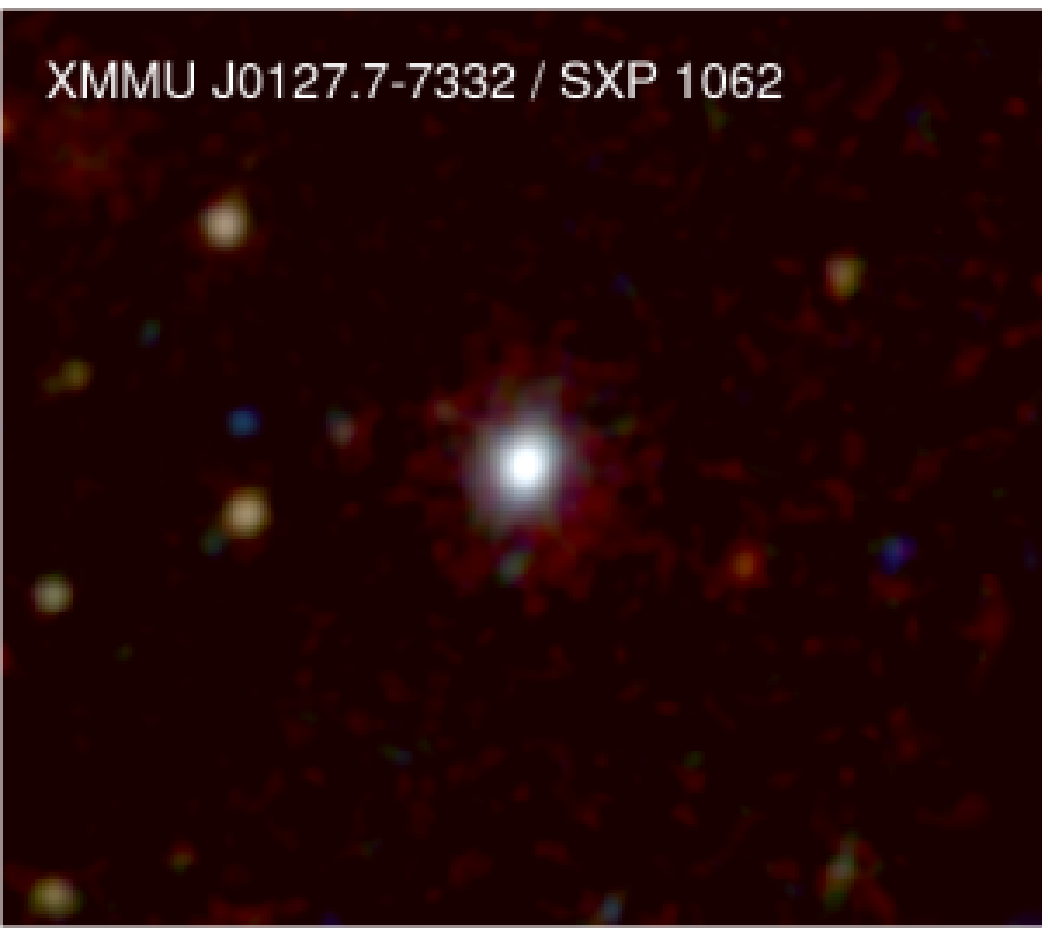}}
  }
  \caption{Supernova remnants in the SMC as seen by EPIC. 
           Colours represent the same energy bands as in Fig.~\ref{fig-rgbima} and 
           the panels are numbered according to Fig.~\ref{fig-rgbima}.
           To more clearly reveal faint diffuse emission, all images were detector background-subtracted
           and both exposure- and vignetting-corrected and are shown with the scale indicated in the upper right. 
           For a description of the markings see Sects.~\ref{sect-snr} (SNRs) and \ref{sect-gcl} (galaxy clusters).
           The bottom right panel shows the only known SMC SNR \citep{2012A&A...537L...1H} not covered by our survey. 
  }
  \label{fig-snrima}
\end{figure*}

In general, SNRs exhibit thermal X-ray spectra with emission mainly below 2 keV. 
Their colours in the RGB images range from red via orange and yellow 
to light green (see Fig.~\ref{fig-snrima}). Remarkable examples are the SNRs in the emission nebula N19 in the south west 
(Fig.~\ref{fig-snrima}, panel 2) where the colours allow us to confine the different remnants. 
Most of the bright SNRs were observed by \xmm\ in the first years of the mission and investigated by \citet{2004A&A...421.1031V}. 
The spectral analyses in their study yielded typical temperatures between 0.3 keV and 2 keV. 
For three fainter remnants with low surface brightness, \citet{2008A&A...485...63F} found lower temperatures of around 0.18 keV, indicating older ages. 

In Fig.~\ref{fig-snrima}, we mark with full circles the position and extent of the 23 SMC SNRs listed in the compilation 
of \defcitealias{2010MNRAS.407.1301B}{B10}\citet[][ hereafter \citetalias{2010MNRAS.407.1301B}]{2010MNRAS.407.1301B}. 
Only three remnants are not seen in our X-ray images: DEM\,S130, N\,S21, and N83C, down to a surface brightness limit of 
$\sim$\oergcm{-14} arcmin$^{-2}$. 
In particular, DEM\,S130 is located in the region with the highest cumulative exposure time near 1E0102.2-7219 without any indication of 
enhanced X-ray emission in our EPIC images (Fig.~\ref{fig-snrima}, panel 1).
At the position of IKT\,7 (Fig.~\ref{fig-snrima}, panel 7), a hard X-ray source is detected, which was identified with the 
172~s Be/X-ray pulsar AX\,J0051.6-7311 \citep{2000PASJ...52L..37Y} by \citet{2004A&A...414..667H}. 
No significant extended emission is seen in our data around IKT\,7. Since IKT\,7 was proposed as an SNR candidate based 
only on hardness ratios from \einstein\ IPC data \citep{1983IAUS..101..535I} and it is not detected in the radio and optical
bands\footnote{Magellanic Cloud Supernova Remnant Database: http://www.mcsnr.org/Default.aspx}, 
we conclude that IKT\,7 is not an SNR.

Several SNRs show a larger extent in the EPIC images than given in \citetalias{2010MNRAS.407.1301B}: 
N\,S19, IKT\,5, B0050$-$728, DEM\,S128, and IKT\,21 (Fig.~\ref{fig-snrima}, thin dashed lines). 
There is no obvious excess X-ray emission on top of N\,S19 (white circle in Fig.~\ref{fig-snrima}, panel 2), 
but the optical SNR might be part of a larger remnant with an elliptical shape 
(4\arcmin\ $\times$ 2.8\arcmin, dashed ellipse in Fig.~\ref{fig-snrima}) or completely unrelated.
IKT\,5 (Fig.~\ref{fig-snrima}, panel 2) shows a more elliptical (peanut) shape extending further to the south.
Similarly, B0050$-$728 (Fig.~\ref{fig-snrima}, panel 4), the largest SNR listed by \citetalias{2010MNRAS.407.1301B}, is either only part of an 
even larger SNR with a diameter of 7\arcmin, or one of a pair of close SNRs with similar temperatures. 
DEM\,S128 (Fig.~\ref{fig-snrima}, panel 1) covers an elliptical area (3.6\arcmin\ $\times$ 2.8\arcmin) more 
than twice that given by \citetalias{2010MNRAS.407.1301B}. Finally, faint X-ray emission is seen around IKT\,21 (Fig.~\ref{fig-snrima}, panel 1) 
within a circular region of diameter 5.2\arcmin, which is more than a factor of five larger than given by \citetalias{2010MNRAS.407.1301B}. 
In Table~\ref{tab-snr}, we summarise the SNRs and new candidates with large extents and list their central coordinates and sizes as seen by \xmm. 

The EPIC images reveal three new candidate SNRs with low surface brightnesses and low temperatures (thick dashed lines in panels 2 and 3 in 
Fig.~\ref{fig-snrima}, marked with their XMM-names). The most compact of these is located east of B0045$-$733 and has a circular shape 
(1.5\arcmin\ diameter). We designate this SNR candidate XMMU\,J0049.0$-$7306.  
Two larger SNR candidates, XMMU\,J0056.5$-$7208 and XMMU\,J0057.7$-$7213, are visible west of IKT\,18, one circularly shaped 
(diameter 3.4\arcmin, but see discussion in Sect.~3) and the other more elliptical (4.8\arcmin\ $\times$ 3\arcmin). 
At the northern end of XMMU\,J0057.7$-$7213, \citet{2005MNRAS.364..217F} list a radio SNR (ATCA source No. 345) that 
partially overlaps with the X-ray emission region. It is again unclear whether we see two remnants or one with different morphologies in 
different wave bands.
In general, our findings increase the number of larger SNRs and, in particular, we extend the size distribution of SMC remnants to larger values.
A more detailed analysis of the SMC SNRs is in progress and will be published elsewhere.

\addtocounter{table}{1} 
\begin{table*}
\caption[]{Supernova remnants and candidates seen with large extents in the EPIC images.}
\label{tab-snr}
\centering
\begin{tabular}{llccc}
\hline\hline\noalign{\smallskip}
\multicolumn{1}{l}{Object name} &
\multicolumn{1}{l}{Fig.} &
\multicolumn{2}{c}{Central Coordinates (J2000)}&
\multicolumn{1}{c}{Extent\tablefootmark{a}} \\
\multicolumn{1}{l}{} &
\multicolumn{1}{l}{} &
\multicolumn{1}{c}{R.A.}&
\multicolumn{1}{c}{Dec.}&
\multicolumn{1}{c}{(pc)} \\

\noalign{\smallskip}\hline\noalign{\smallskip}
B0050$-$728          & \ref{fig-snrima} (4)  & 00:52:56.6 & -72:36:24 & 61 \\
N\,S19 ?             & \ref{fig-snrima} (2)  & 00:48:22.8 & -73:07:55 & 24/35/110\degr \\
IKT\,21              & \ref{fig-snrima} (1)  & 01:03:21.1 & -72:08:37 & 45 \\
DEM S128             & \ref{fig-snrima} (1)  & 01:05:27.0 & -72:10:38 & 24/31/70\degr \\
XMMU\,J0049.0$-$7306 & \ref{fig-snrima} (2)  & 00:49:00.2 & -73:06:17 & 13 \\
XMMU\,J0056.5$-$7208 & \ref{fig-snrima} (3)  & 00:56:30.2 & -72:08:12 & 30 \\
XMMU\,J0057.7$-$7213 & \ref{fig-snrima} (3)  & 00:57:46.0 & -72:13:04 & 26/42/110\degr \\

\noalign{\smallskip}\hline
\end{tabular}
\tablefoot{\tablefoottext{a}{Radius of circle or semi-minor and -major axes with orientation angle for ellipses (from west to east).}}
\end{table*}

\begin{figure}
   \resizebox{0.97\hsize}{!}{\includegraphics[clip=,angle=-90]{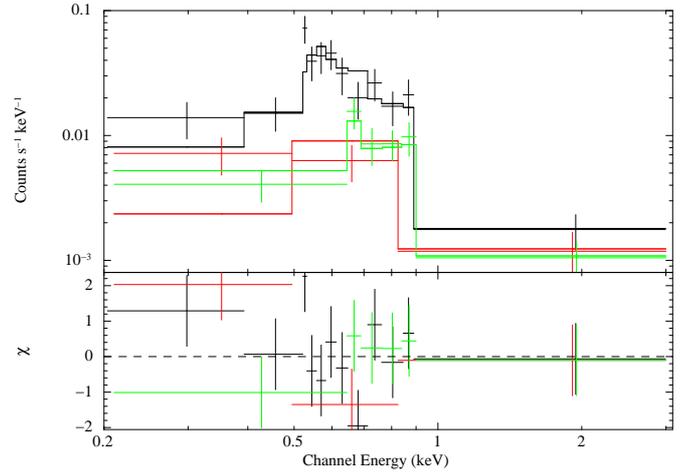}}
   \caption{EPIC pn (black) and MOS (red and green) spectra of the candidate SNR XMMU\,J0056.5$-$7208 extracted from observation 0601210601 
   (with net exposures of 29.6\,ks and 34.1\,ks, respectively). 
   Top: data with best-fit Sedov model as histogram, bottom: residuals in units of sigma.}
   \label{fig-snrspec}
\end{figure}

The X-ray spectra of the new candidate SNRs are soft, indicating low temperatures. As an example, we show the EPIC spectra of 
XMMU\,J0056.5$-$7208 (Fig.~\ref{fig-snrspec}). 
We selected single-pixel (PATTERN=0) events for EPIC pn and all valid events (PATTERN 0-12) for MOS spectra 
(with FLAG=0 for both instruments), 
resulting in 550, 230, and 220 net counts for the pn, MOS1, and MOS2 spectra, respectively.
Owing to the large extraction area (circle with radius 96\arcsec), the background ($\sim$70\%) dominates the total spectrum and therefore, 
we binned to a signal-to-noise ratio of at least 3 per bin, which also ensured a minimum of 25 source counts per bin.
To model the spectrum, we used the Sedov (vsedov in XSPEC) model \citep[][]{2001ApJ...548..820B} based on NEI-version 2.0.
Metal abundances were fixed to 0.2 solar, as is typical in the SMC \citep{1992ApJ...384..508R}.
To reduce the number of free model parameters in the fit, we forced the mean shock temperature and the electron temperature immediately behind the shock 
front to be the same.
The absorption column density was divided into two parts, a fixed value of 6\hcm{20} with solar abundances for the Galactic foreground
\citep{1990ARAA...28..215D}, and a column density, which was allowed to vary in the fit, to account for the absorption in the SMC 
\citep[where also the abundances where set to 0.2 and limited to a maximum of 3\hcm{21}, the total SMC absorption in the direction of the SNR; ][]{1999MNRAS.302..417S}.
Under these assumptions, we derived an upper limit to the SMC column of 3\hcm{21} and a shock temperature of 0.5 keV, 
which is also poorly constrained with an upper limit of 2.3 keV. 
Also for the ionisation time-scale, only an upper limit of 1.5$\times$10$^{13}$ s cm$^{-3}$ could be derived. 
The formal best-fit model has a reduced $\chi^2$ of 1.28 for 16 degrees of freedom.
The total observed flux in the 0.2-2.0 keV energy band was derived to be 5\ergcm{-14}, which translates into a surface brightness of 2.8\ergcm{-14} arcmin$^{-2}$.
Using the radius-temperature relation t$_{\rm y}$\,=\,3.8\expo{2}\,R$_{\rm pc}$(kT)$_{\rm keV}^{-1/2}$ \citep[e.g. ][]{2005ChJAA...5..165X}, 
we estimate the dynamical age of the SNR to be $\sim$16 kyr.
However, we stress that the low statistical quality of the X-ray spectra, which have a high background, 
does not allow us to differentiate between the various models generally used to explain 
SNR spectra and more X-ray data is required for all the faint SNRs in the SMC to derive more reliable parameters.

\begin{figure*}
   \resizebox{0.96\hsize}{!}{\includegraphics[clip=]{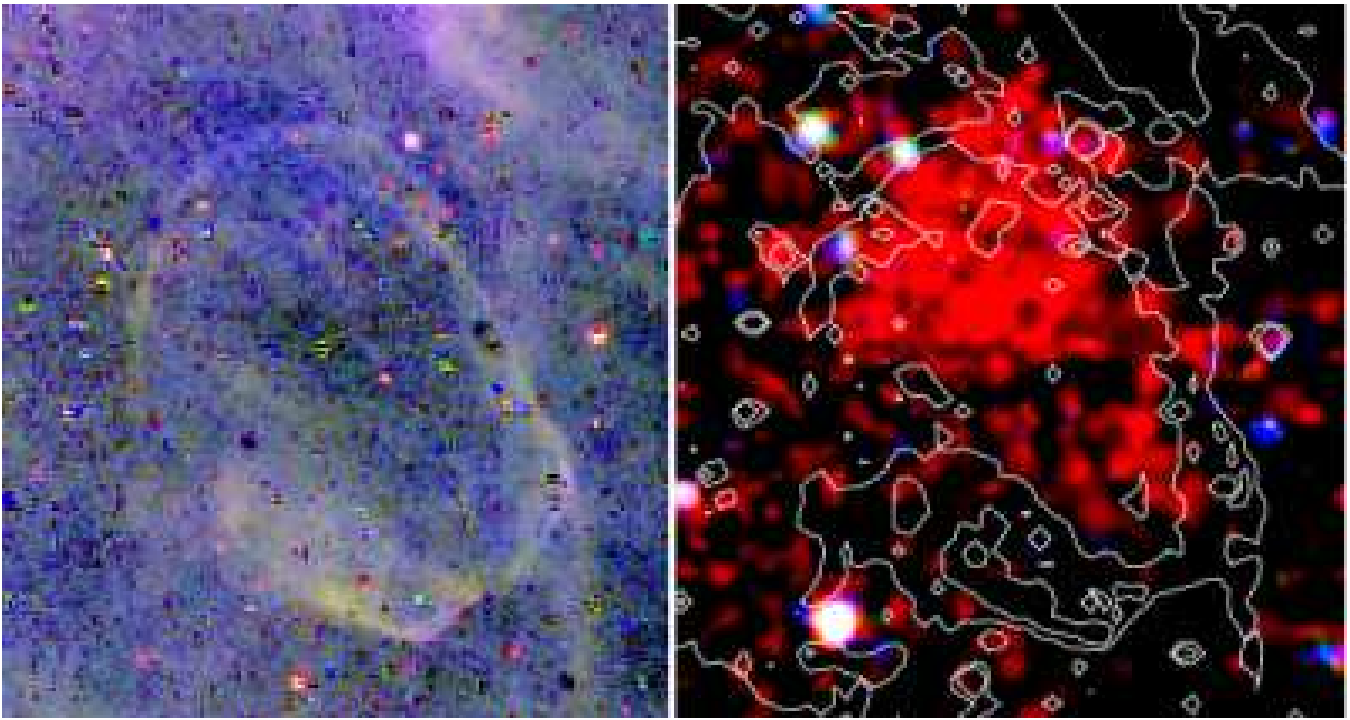}}
   \caption{The region around XMMU\,J0056.5$-$7208. 
   Left: Zoom-in of the MCELS image from Fig.~\ref{fig-MCELS}. 
   Right: Zoom-in of the X-ray image from Fig.~\ref{fig-snrima}, panel 3 with enhanced contrast. 
          The contours mark the summed \Halpha, [\SII], and [\OIII] intensity at 100 and 150\ergcm{-17}.}
   \label{fig-snrmcels}
\end{figure*}

We examined the MCELS images in the regions around
the new SNR candidates. Only for XMMU\,J0056.5$-$7208 (Fig.~\ref{fig-snrmcels}) is a shell structure visible, mainly in \Halpha\ and [\SII].
The structure is of elliptical shape, but could also be composed of two merging circular shells. The EPIC image with enhanced contrast in 
Fig.~\ref{fig-snrmcels} appears to also reveal weak X-ray emission in the southern part where the optical emission is enhanced. 
In the brighter parts in the south, the shell has a ratio of the [\SII] to \Halpha\ intensity of around 0.4. 
This is at the lower limit of what is generally accepted for an SNR \citep[typically $>$ 0.4;][]{1985ApJ...292...29F}, but given the low 
elemental abundance in the SMC, lower ratios are expected and also seen for other well-established SMC SNRs
from the full MCELS image and optical spectroscopy \citep[c.f.][]{2007MNRAS.376.1793P}.

\subsection{Clusters of galaxies}
\label{sect-gcl}

\begin{figure}
  \parbox[b]{\hsize}{
    \resizebox{0.5\hsize}{!}{\includegraphics[clip=]{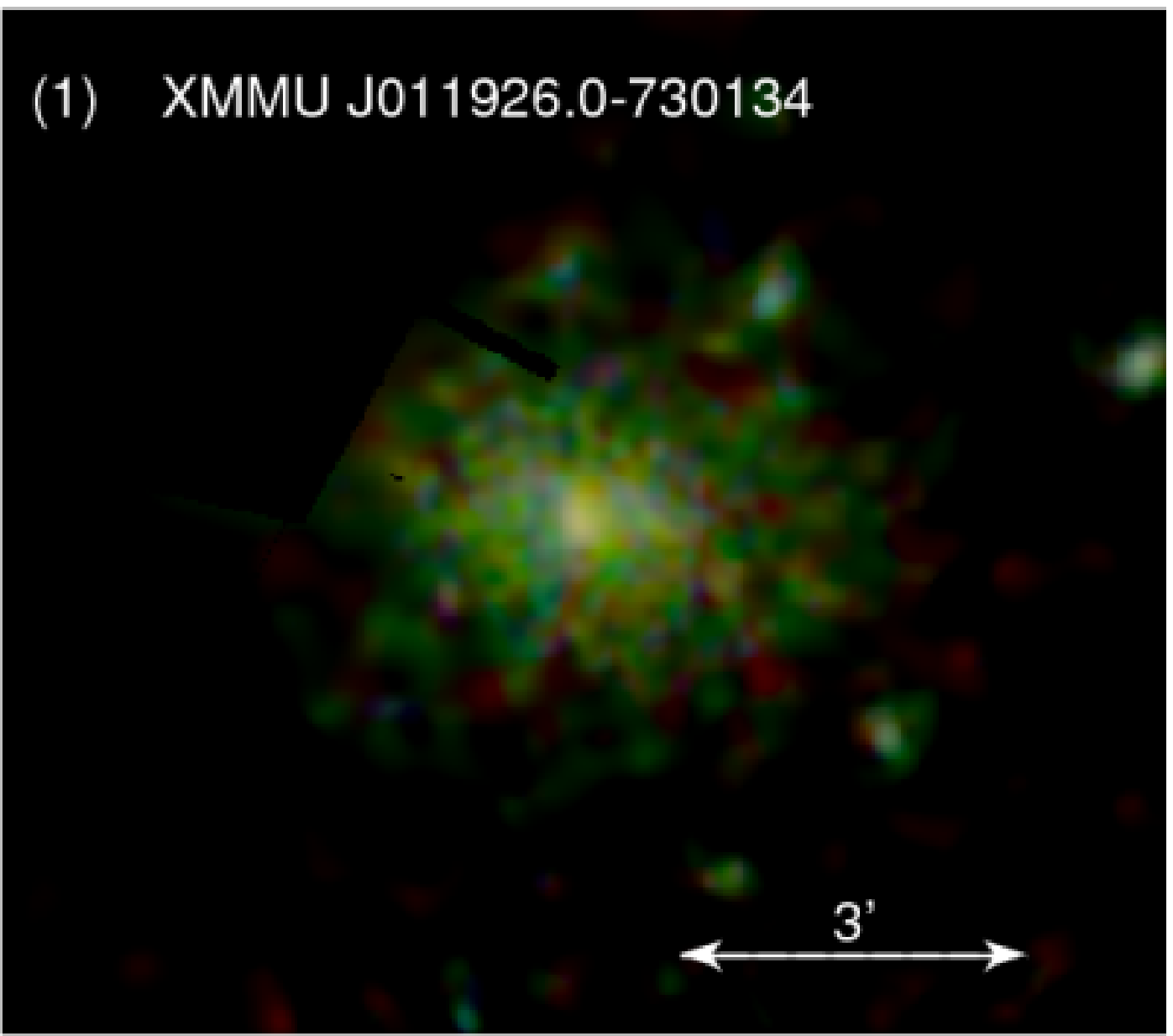}}
    \resizebox{0.5\hsize}{!}{\includegraphics[clip=]{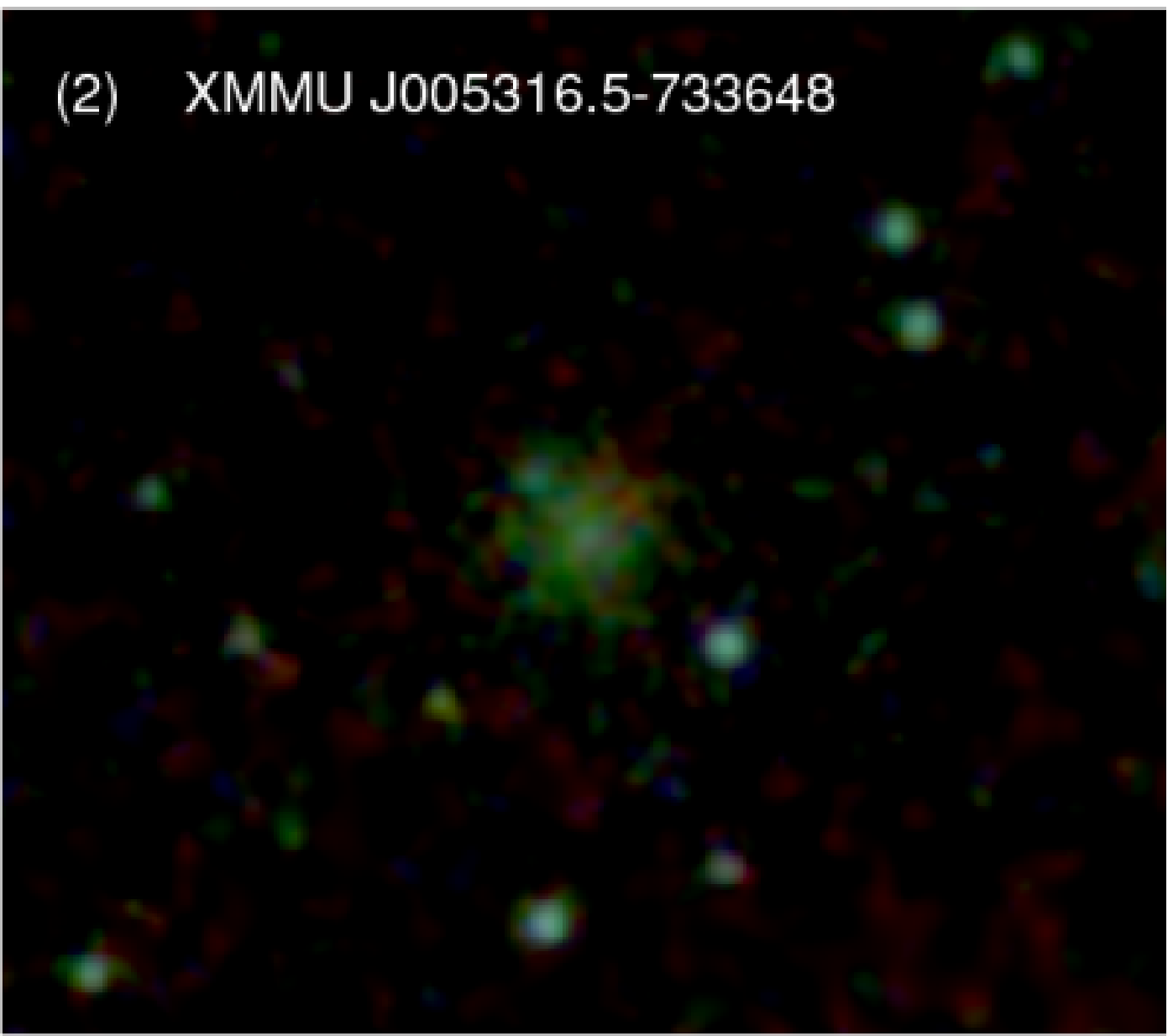}}
    \\ \noindent
    \resizebox{0.5\hsize}{!}{\includegraphics[clip=]{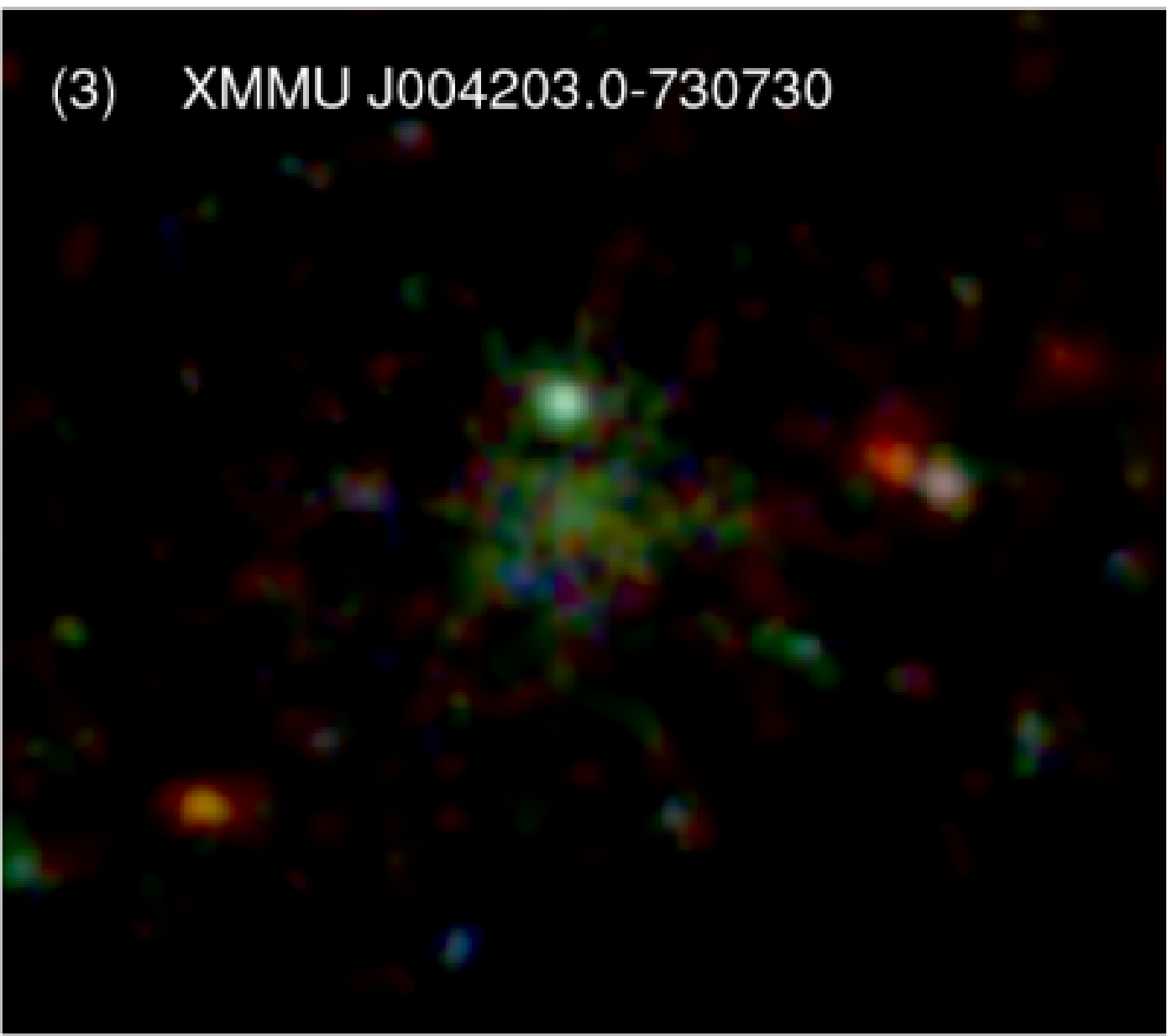}}
    \resizebox{0.5\hsize}{!}{\includegraphics[clip=]{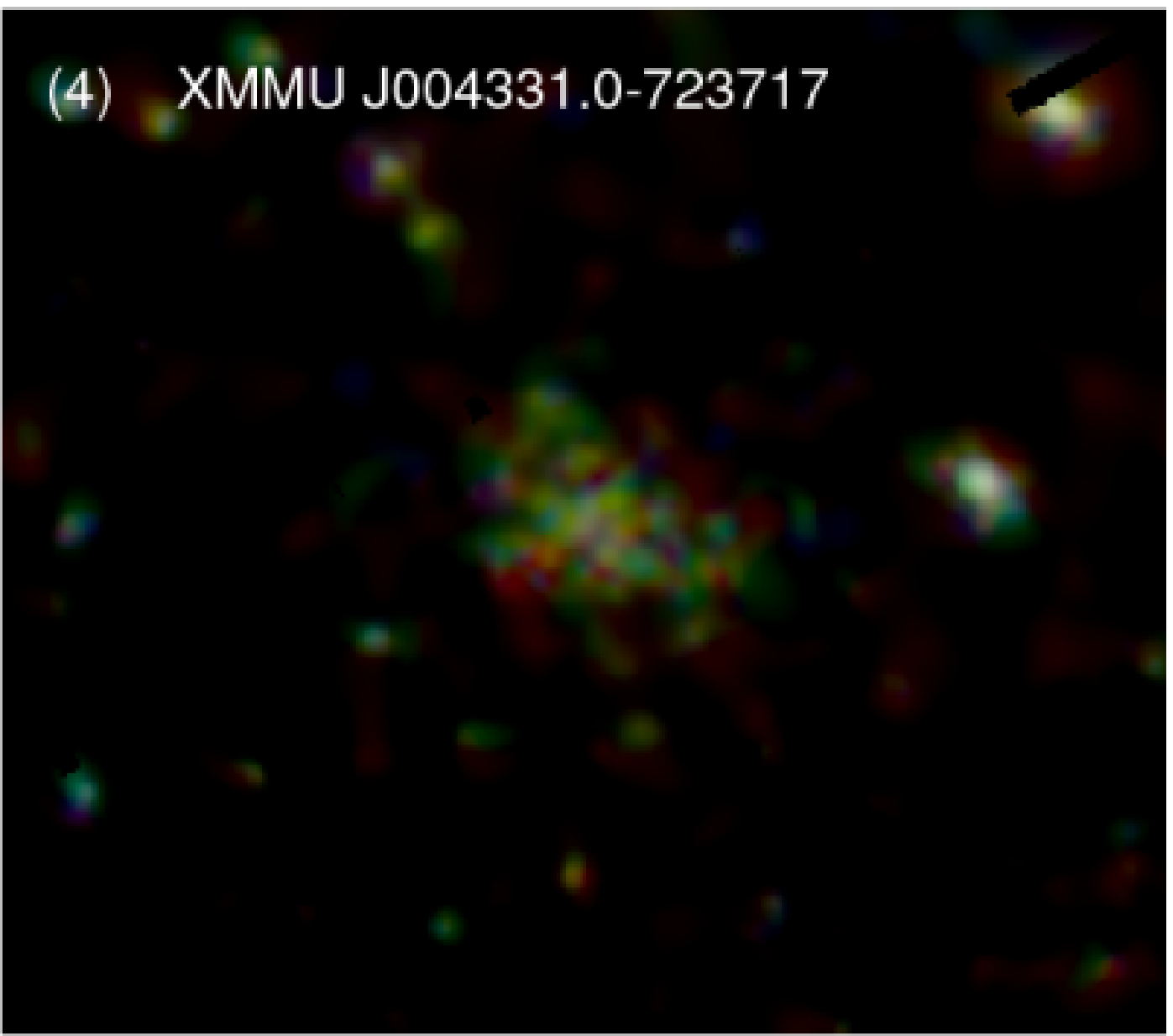}}
    \\ \noindent
    \resizebox{0.5\hsize}{!}{\includegraphics[clip=]{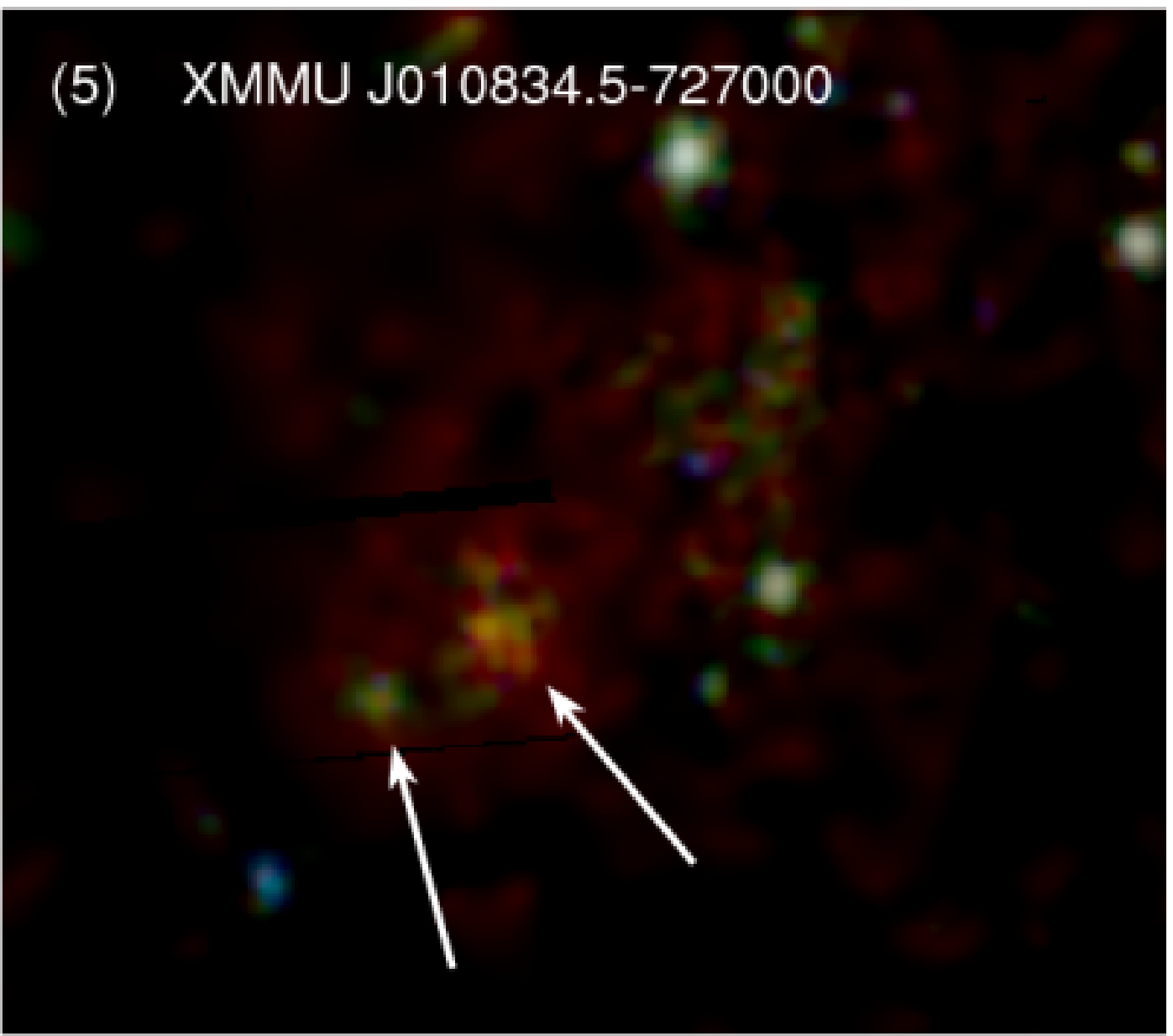}}
  }
  \caption{Candidates for clusters of galaxies seen in the SMC field. Two sources from the catalogue identified with galaxies are marked with arrows. 
           Presentation as in Fig.~\ref{fig-snrima}.
  }
  \label{fig-gclima}
\end{figure}

The EPIC images reveal other extended sources, which appear in darker green in the RGB images (Fig.~\ref{fig-rgbima}) indicating that they have spectra 
harder than those of supernova remnants. These sources are not concentrated in the SMC bar where most SNRs are found, which 
suggests that they are most likely clusters of galaxies behind the SMC. 
In our galaxy cluster classification, we also include groups of galaxies that are characterised by X-ray emitting gas with 
somewhat lower temperatures than more massive clusters \citep{2000ARA&A..38..289M}. 
In rare cases, giant \HII\ regions can also show relatively hard X-ray spectra \citep[e.g. IC131 in M\,33, ][]{2009ApJ...707.1361T}, 
but these can be identified as such by looking for emission at other wavelengths such as \Halpha. 
Five fields, including the most prominent clusters of galaxies, are marked with green boxes in Fig.~\ref{fig-rgbima} and zoomed 
images are shown in Fig.~\ref{fig-gclima}. 
Three more can be seen in the SNR images in Fig.~\ref{fig-snrima}, which are marked with arrows in panels 1 and 11. A summary of the candidate 
clusters of galaxies can be found in Table~\ref{tab-gcl}.

\begin{figure}
   \resizebox{0.97\hsize}{!}{\includegraphics[clip=,angle=-90]{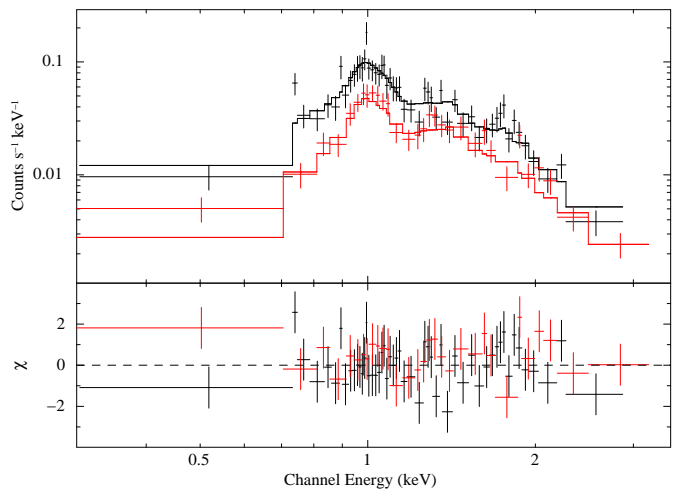}}
   \caption{EPIC pn (black) and MOS2 (red) spectra of the candidate galaxy cluster \gclxmm, fitted with an absorbed thin plasma emission (mekal) model. 
   Net exposures are 27.4 ks and 32.9 ks, respectively.
   Top: data with best-fit model as histograms. Bottom: residuals in units of sigma.}
   \label{fig-gclspec}
 \end{figure}

The X-ray brightest and most extended candidate cluster is XMMU\,J011926.0-730134, which is located in the eastern wing of the SMC (north east of SMC\,X-1). 
The EPIC pn and MOS2 instruments (no MOS1 data is available as the source was located on MOS1 CCD\,6, which was no longer operating at the time of 
observation 0601212301) collected $\sim$3100 counts within a circular area of radius 2\arcmin, which was sufficient for our spectral analysis.
Spectra were extracted and binned in the same way as described in Sect.~\ref{sect-snr}.
We modelled the spectra with plasma emission attenuated by three absorption components (phabs*vphabs*vphabs*vmekal in XSPEC).
The first accounts for the absorption in the Milky Way with a column density fixed at 4.5\hcm{20} \citep{1990ARAA...28..215D}
and assuming solar abundances \citep{2000ApJ...542..914W}, 
the second models the absorption through the SMC with column density fixed at 5.2\hcm{21} \citep{1999MNRAS.302..417S} and SMC abundances, 
and the third, with a column density that was allowed to vary in the fit, is responsible for the remaining absorption to the galaxy cluster 
(here we assume an abundance of 0.5 solar).
This model results in an acceptable fit with a reduced $\chi^2$ of 1.01 for 79 degrees of freedom (Fig.~\ref{fig-gclspec}). 
The best-fit value of the absorbing column density is 1.9$\pm$0.9\hcm{21}, of the temperature 1.5$\pm$0.1 keV and of the redshift z=0.052$\pm$0.013 
(errors denote 90\% confidence ranges for one parameter of interest).
Two known galaxies are listed in the 6dF galaxy survey \citep{2009MNRAS.399..683J} within an angular distance of 1\arcmin\ from the cluster centre,
g0119275-730130 with a redshift of 0.0656 and g0119251-730217 with z=0.0677. Both redshifts are within the uncertainty range for the 
X-ray determined redshift, supporting the cluster nature of the X-ray emission.

\begin{table}
\caption[]{Candidates for galaxy clusters.}
\begin{center}
\begin{tabular}{lll}
\hline\hline\noalign{\smallskip}
\multicolumn{1}{l}{Object name} &
\multicolumn{1}{l}{Fig.} &
\multicolumn{1}{l}{Comment} \\
\multicolumn{1}{l}{XMMU} &
\multicolumn{1}{l}{} &
\multicolumn{1}{l}{} \\

\noalign{\smallskip}\hline\noalign{\smallskip}
J004203.0-730730 & \ref{fig-gclima} (3)  & radio jets \citep{Sturm_agn} \\
J004331.0-723717 & \ref{fig-gclima} (4)  & \\
J005107.0-733923 & $-$                   & [SHP] 708/709 \\
J005316.5-733648 & \ref{fig-gclima} (2)  & \\
J005821.0-720034 & $-$                   & \\
J005915.0-723640 & $-$                   & near bright source \\
J010408.5-724401 & \ref{fig-snrima} (11) & [SHP12]\,571 \\
J010414.0-720813 & \ref{fig-snrima} (1)  & south of 1E0102.2-7219 \\
J010454.5-721022 & \ref{fig-snrima} (1)  & AGN in centre \citep{Sturm_agn} \\
J010834.5-727000 & \ref{fig-gclima} (5)  & 2 galaxies + cluster? \\
J011631.0-725803 & $-$                   & [SHP] 2917 \\
J\gclxco         & \ref{fig-gclima} (1)  & z=0.07; 2 galaxies in 6dF survey \\

\noalign{\smallskip}\hline
\end{tabular}
\end{center}
\label{tab-gcl}
\end{table}

A comparison with deep radio images of the SMC \citep[][ and references therein]{Sturm_agn} revealed radio emission with jet-like structures at the position 
of XMMU\,J004203.0-730730. Their morphology suggests that they arise from two independent galaxies each possessing a single jet or from one galaxy 
moving through the intra-cluster medium and leaving the jets behind. A radio point source was also found in the centre of 
XMMU\,J010454.5-721022, which is most likely associated with an AGN. The host galaxy likely belongs to the cluster, although it 
cannot be excluded that it is located behind or in front of the cluster. Additional, more distant galaxy clusters that appear as  
sources with small extents are included in the point source catalogue of \citetalias{Sturm_cat}.

A region with complex extended emission is shown in Fig.~\ref{fig-gclima}, panel 5. In the south-east, the source catalogue contains 
two sources, which are classified as galaxies (1711 and 1726, indicated by arrows). Source 1711 is identified with the galaxy 
2MASX\,J01090126-7229058 with an extent of 0.6\arcmin\ $\times$ 0.7\arcmin\ \citep{2006AJ....131.1163S}. Further to the north-west, 
weak extended emission might originate from hot cluster gas, but it remains unclear whether the galaxies belong to this cluster or they are unrelated.


\section{Discussion and conclusions}

We have performed a deep X-ray survey of the SMC using the EPIC instruments on board \xmm\ between May 2009 and March 2010, which, together with 
archival observations, covers the bar and eastern wing of the galaxy. While a detailed analysis of the population of discrete X-ray sources 
(unresolved and with small extents) is described in \citetalias{Sturm_cat}, we present here first mosaic images of the main body of 
the SMC. We study the morphology of more extended X-ray emission from supernova remnants and galaxy clusters. Several galaxy clusters 
can easily be identified in our EPIC images by their extended appearance and X-ray colours. 

We have found that several of the known SNRs listed with their spatial extent in the catalogue of \citetalias{2010MNRAS.407.1301B} 
appear larger in the X-ray images.
This is particularly true for large SNRs with low temperatures (suggesting old remnants) and 
low surface brightnesses, which can be better mapped with the high sensitivity of \xmm\ in soft X-rays.
However, a larger extent could also be caused by the presence of more than one SNR merging into larger bubbles. In these cases, the 
consistent X-ray colours require the neighbouring remnants, which could have formed in star clusters, to have similar ages. 

We propose three new SNRs with large extents, low surface brightnesses, and soft X-ray spectra (labelled with their XMM names in 
Fig.~\ref{fig-snrima}). The high sensitivity of the EPIC instruments at low energies allowed us to detect their faint and soft emission with a 
typical surface brightness of \oergcm{-14} arcmin$^{-1}$ (0.2-2.0 keV). Only one of the three remnants (XMMU\,J0056.5$-$7208) is seen 
in the MCELS images in the 
\Halpha, [\SII], and [\OIII] emission lines. Shell-like optical filaments suggest that we see either one SNR extending further to the south-west of 
the bulk of the X-ray emission or two merging SNR shells with the southern one being fainter in X-rays but brighter in the optical than the northern shell.
Our survey covers all other known SNRs in the SMC, except the recently discovered faint SNR far out in the eastern wing, which is located outside 
our survey area \citep{2012A&A...537L...1H} and also exhibits a low surface brightness and low temperature, indicative of older SNRs.
For completeness, the image of this remnant (designated XMMU\,J0127.7$-$7332), which harbours the Be/X-ray binary pulsar SXP\,1062 at its centre
is included in the image gallery of Fig.~\ref{fig-snrima}.
Overall, our studies of the X-ray morphology of the SNRs in the SMC extend their size distribution to larger values. 
In particular for B0050$-$728, the largest SNR known in the SMC, we have found an X-ray diameter as large as 7', 
which corresponds to 122 pc at a distance of 60 kpc. In the MCs, this is only exceeded by the LMC SNR DEM\,L203 \citepalias{2010MNRAS.407.1301B}.

\begin{acknowledgements}
The \xmm\ project is supported by the Bundesministerium f\"ur Wirtschaft und 
Technologie/Deutsches Zentrum f\"ur Luft- und Raumfahrt (BMWI/DLR, FKZ 50 OX 0001)
and the Max-Planck Society. 
The Magellanic Clouds Emission Line Survey (MCELS) data were provided by R.~C. Smith, P.~F. Winkler, and S.~D. Points. The MCELS project has been 
supported in part by NSF grants AST-9540747 and AST-0307613, and through the generous support of the Dean B. McLaughlin Fund at the University of 
Michigan, a bequest from the family of Dr. Dean B. McLaughlin in memory of his lasting impact on Astronomy. The National Optical Astronomy 
Observatory is operated by the Association of Universities for Research in Astronomy Inc. (AURA), under a cooperative agreement with the 
National Science Foundation.
R. Sturm acknowledges support from the BMWI/DLR, FKZ 50 OR 0907.
N. La Palombara, S.L. Mereghetti and A. Tiengo acknowledge financial contributions by the Italian Space Agency through ASI/INAF
agreements I/009/10/0 and I/032/10/0 for data analysis and \xmm\ operations, respectively.
\end{acknowledgements}

\bibliographystyle{aa}
\bibliography{general}

\begin{thebibliography}{61}
\expandafter\ifx\csname natexlab\endcsname\relax\def\natexlab#1{#1}\fi

\bibitem[{{Antoniou} {et~al.}(2010){Antoniou}, {Zezas}, {Hatzidimitriou}, \&
  {Kalogera}}]{2010ApJ...716L.140A}
{Antoniou}, V., {Zezas}, A., {Hatzidimitriou}, D., \& {Kalogera}, V. 2010,
  \apjl, 716, L140

\bibitem[{{Badenes} {et~al.}(2010){Badenes}, {Maoz}, \&
  {Draine}}]{2010MNRAS.407.1301B}
{Badenes}, C., {Maoz}, D., \& {Draine}, B.~T. 2010, \mnras, 407, 1301 (B10)

\bibitem[{{Bauer} {et~al.}(2008){Bauer}, {Pietsch}, {Trinchieri},
  {Breitschwerdt}, {Ehle}, {Freyberg}, \& {Read}}]{2008A&A...489.1029B}
{Bauer}, M., {Pietsch}, W., {Trinchieri}, G., {et~al.} 2008, \aap, 489, 1029

\bibitem[{{Borkowski} {et~al.}(2001){Borkowski}, {Lyerly}, \&
  {Reynolds}}]{2001ApJ...548..820B}
{Borkowski}, K.~J., {Lyerly}, W.~J., \& {Reynolds}, S.~P. 2001, \apj, 548, 820

\bibitem[{{Coe} {et~al.}(2012){Coe}, {Haberl}, {Sturm}, {Bartlett},
  {Hatzidimitriou}, {Townsend}, {Udalski}, {Mereghetti}, \&
  {Filipovi{\'c}}}]{2012MNRAS.424..282C}
{Coe}, M.~J., {Haberl}, F., {Sturm}, R., {et~al.} 2012, \mnras, 424, 282

\bibitem[{{Coe} {et~al.}(2011){Coe}, {Haberl}, {Sturm}, {Pietsch}, {Townsend},
  {Bartlett}, {Filipovic}, {Udalski}, {Corbet}, {Tiengo}, {Ehle}, {Payne}, \&
  {Burton}}]{2011MNRAS.414.3281C}
{Coe}, M.~J., {Haberl}, F., {Sturm}, R., {et~al.} 2011, \mnras, 414, 3281

\bibitem[{{Dickey} \& {Lockman}(1990)}]{1990ARAA...28..215D}
{Dickey}, J.~M. \& {Lockman}, F.~J. 1990, Ann. Rev. Astron. Astrophys., 28, 215

\bibitem[{{Fesen} {et~al.}(1985){Fesen}, {Blair}, \&
  {Kirshner}}]{1985ApJ...292...29F}
{Fesen}, R.~A., {Blair}, W.~P., \& {Kirshner}, R.~P. 1985, \apj, 292, 29

\bibitem[{{Filipovi{\'c}} {et~al.}(2008){Filipovi{\'c}}, {Haberl}, {Winkler},
  {Pietsch}, {Payne}, {Crawford}, {de Horta}, {Stootman}, \&
  {Reaser}}]{2008A&A...485...63F}
{Filipovi{\'c}}, M.~D., {Haberl}, F., {Winkler}, P.~F., {et~al.} 2008, \aap,
  485, 63

\bibitem[{{Filipovi{\'c}} {et~al.}(2005){Filipovi{\'c}}, {Payne}, {Reid},
  {Danforth}, {Staveley-Smith}, {Jones}, \& {White}}]{2005MNRAS.364..217F}
{Filipovi{\'c}}, M.~D., {Payne}, J.~L., {Reid}, W., {et~al.} 2005, \mnras, 364,
  217

\bibitem[{{Greiner}(1996)}]{1996LNP...472Q.299G}
{Greiner}, J. 1996, in Lecture Notes in Physics, Berlin Springer Verlag, Vol.
  472, Supersoft X-Ray Sources, ed. J.~{Greiner}, 299--337

\bibitem[{{Haberl} {et~al.}(2000){Haberl}, {Filipovi{\'c}}, {Pietsch}, \&
  {Kahabka}}]{2000A&AS..142...41H}
{Haberl}, F., {Filipovi{\'c}}, M.~D., {Pietsch}, W., \& {Kahabka}, P. 2000,
  \aaps, 142, 41

\bibitem[{{Haberl} \& {Pietsch}(1999)}]{1999A&AS..139..277H}
{Haberl}, F. \& {Pietsch}, W. 1999, \aaps, 139, 277

\bibitem[{{Haberl} \& {Pietsch}(2004)}]{2004A&A...414..667H}
{Haberl}, F. \& {Pietsch}, W. 2004, \aap, 414, 667

\bibitem[{{Haberl} {et~al.}(2012){Haberl}, {Sturm}, {Filipovi{\'c}}, {Pietsch},
  \& {Crawford}}]{2012A&A...537L...1H}
{Haberl}, F., {Sturm}, R., {Filipovi{\'c}}, M.~D., {Pietsch}, W., \&
  {Crawford}, E.~J. 2012, \aap, 537, L1

\bibitem[{{Henze} {et~al.}(2010){Henze}, {Pietsch}, {Haberl}, {Hernanz},
  {Sala}, {Della Valle}, {Hatzidimitriou}, {Rau}, {Hartmann}, {Greiner},
  {Burwitz}, \& {Fliri}}]{2010A&A...523A..89H}
{Henze}, M., {Pietsch}, W., {Haberl}, F., {et~al.} 2010, \aap, 523, A89

\bibitem[{{Henze} {et~al.}(2011){Henze}, {Pietsch}, {Haberl}, {Hernanz},
  {Sala}, {Hatzidimitriou}, {Della Valle}, {Rau}, {Hartmann}, \&
  {Burwitz}}]{2011A&A...533A..52H}
{Henze}, M., {Pietsch}, W., {Haberl}, F., {et~al.} 2011, \aap, 533, A52

\bibitem[{{Hilditch} {et~al.}(2005){Hilditch}, {Howarth}, \&
  {Harries}}]{2005MNRAS.357..304H}
{Hilditch}, R.~W., {Howarth}, I.~D., \& {Harries}, T.~J. 2005, \mnras, 357, 304

\bibitem[{{Inoue} {et~al.}(1983){Inoue}, {Koyama}, \&
  {Tanaka}}]{1983IAUS..101..535I}
{Inoue}, H., {Koyama}, K., \& {Tanaka}, Y. 1983, in IAU Symposium, Vol. 101,
  Supernova Remnants and their X-ray Emission, ed. J.~{Danziger} \&
  P.~{Gorenstein}, 535--540

\bibitem[{{Jones} {et~al.}(2009){Jones}, {Read}, {Saunders}, {Colless},
  {Jarrett}, {Parker}, {Fairall}, {Mauch}, {Sadler}, {Watson}, {Burton},
  {Campbell}, {Cass}, {Croom}, {Dawe}, {Fiegert}, {Frankcombe}, {Hartley},
  {Huchra}, {James}, {Kirby}, {Lahav}, {Lucey}, {Mamon}, {Moore}, {Peterson},
  {Prior}, {Proust}, {Russell}, {Safouris}, {Wakamatsu}, {Westra}, \&
  {Williams}}]{2009MNRAS.399..683J}
{Jones}, D.~H., {Read}, M.~A., {Saunders}, W., {et~al.} 2009, \mnras, 399, 683

\bibitem[{{Kahabka} {et~al.}(1999){Kahabka}, {Pietsch}, {Filipovi{\'c} }, \&
  {Haberl}}]{1999A&AS..136...81K}
{Kahabka}, P., {Pietsch}, W., {Filipovi{\'c} }, M.~D., \& {Haberl}, F. 1999,
  \aaps, 136, 81

\bibitem[{{Kuntz} \& {Snowden}(2008)}]{2008A&A...478..575K}
{Kuntz}, K.~D. \& {Snowden}, S.~L. 2008, \aap, 478, 575

\bibitem[{{Long} {et~al.}(2010){Long}, {Blair}, {Winkler}, {Becker}, {Gaetz},
  {Ghavamian}, {Helfand}, {Hughes}, {Kirshner}, {Kuntz}, {McNeil}, {Pannuti},
  {Plucinsky}, {Saul}, {T{\"u}llmann}, \& {Williams}}]{2010ApJS..187..495L}
{Long}, K.~S., {Blair}, W.~P., {Winkler}, P.~F., {et~al.} 2010, \apjs, 187, 495

\bibitem[{{Long} {et~al.}(1981){Long}, {Helfand}, \&
  {Grabelsky}}]{1981ApJ...248..925L}
{Long}, K.~S., {Helfand}, D.~J., \& {Grabelsky}, D.~A. 1981, \apj, 248, 925

\bibitem[{{McGowan} {et~al.}(2008){McGowan}, {Coe}, {Schurch}, {McBride},
  {Galache}, {Edge}, {Corbet}, {Laycock}, \& {Buckley}}]{2008MNRAS.383..330M}
{McGowan}, K.~E., {Coe}, M.~J., {Schurch}, M.~P.~E., {et~al.} 2008, \mnras,
  383, 330

\bibitem[{{Mereghetti} {et~al.}(2010){Mereghetti}, {Krachmalnicoff}, {La
  Palombara}, {Tiengo}, {Rauch}, {Haberl}, {Filipovi{\'c}}, \&
  {Sturm}}]{2010A&A...519A..42M}
{Mereghetti}, S., {Krachmalnicoff}, N., {La Palombara}, N., {et~al.} 2010,
  \aap, 519, A42

\bibitem[{{Misanovic} {et~al.}(2006){Misanovic}, {Pietsch}, {Haberl}, {Ehle},
  {Hatzidimitriou}, \& {Trinchieri}}]{2006A&A...448.1247M}
{Misanovic}, Z., {Pietsch}, W., {Haberl}, F., {et~al.} 2006, \aap, 448, 1247

\bibitem[{{Mulchaey}(2000)}]{2000ARA&A..38..289M}
{Mulchaey}, J.~S. 2000, \araa, 38, 289

\bibitem[{{Novara} {et~al.}(2011){Novara}, {La Palombara}, {Mereghetti},
  {Haberl}, {Coe}, {Filipovic}, {Udalski}, {Paizis}, {Pietsch}, {Sturm},
  {Gilfanov}, {Tiengo}, {Payne}, {Smits}, \& {de Horta}}]{2011A&A...532A.153N}
{Novara}, G., {La Palombara}, N., {Mereghetti}, S., {et~al.} 2011, \aap, 532,
  A153

\bibitem[{{Owen} {et~al.}(2011){Owen}, {Filipovi{\'c}}, {Ballet}, {Haberl},
  {Crawford}, {Payne}, {Sturm}, {Pietsch}, {Mereghetti}, {Ehle}, {Tiengo},
  {Coe}, {Hatzidimitriou}, \& {Buckley}}]{2011A&A...530A.132O}
{Owen}, R.~A., {Filipovi{\'c}}, M.~D., {Ballet}, J., {et~al.} 2011, \aap, 530,
  A132

\bibitem[{{Payne} {et~al.}(2007){Payne}, {White}, {Filipovi{\'c}}, \&
  {Pannuti}}]{2007MNRAS.376.1793P}
{Payne}, J.~L., {White}, G.~L., {Filipovi{\'c}}, M.~D., \& {Pannuti}, T.~G.
  2007, \mnras, 376, 1793

\bibitem[{{Pietsch} {et~al.}(2005{\natexlab{a}}){Pietsch}, {Fliri}, {Freyberg},
  {Greiner}, {Haberl}, {Riffeser}, \& {Sala}}]{2005A&A...442..879P}
{Pietsch}, W., {Fliri}, J., {Freyberg}, M.~J., {et~al.} 2005{\natexlab{a}},
  \aap, 442, 879

\bibitem[{{Pietsch} {et~al.}(2005{\natexlab{b}}){Pietsch}, {Freyberg}, \&
  {Haberl}}]{2005A&A...434..483P}
{Pietsch}, W., {Freyberg}, M., \& {Haberl}, F. 2005{\natexlab{b}}, \aap, 434,
  483

\bibitem[{{Pietsch} \& {Haberl}(2005)}]{2005A&A...430L..45P}
{Pietsch}, W. \& {Haberl}, F. 2005, \aap, 430, L45

\bibitem[{{Pietsch} {et~al.}(2007){Pietsch}, {Haberl}, {Sala}, {Stiele},
  {Hornoch}, {Riffeser}, {Fliri}, {Bender}, {B{\"u}hler}, {Burwitz}, {Greiner},
  \& {Seitz}}]{2007A&A...465..375P}
{Pietsch}, W., {Haberl}, F., {Sala}, G., {et~al.} 2007, \aap, 465, 375

\bibitem[{{Pietsch} {et~al.}(2004){Pietsch}, {Misanovic}, {Haberl},
  {Hatzidimitriou}, {Ehle}, \& {Trinchieri}}]{2004A&A...426...11P}
{Pietsch}, W., {Misanovic}, Z., {Haberl}, F., {et~al.} 2004, \aap, 426, 11

\bibitem[{{Russell} \& {Dopita}(1992)}]{1992ApJ...384..508R}
{Russell}, S.~C. \& {Dopita}, M.~A. 1992, \apj, 384, 508

\bibitem[{{Sanduleak} {et~al.}(1978){Sanduleak}, {MacConnell}, \&
  {Philip}}]{1978PASP...90..621S}
{Sanduleak}, N., {MacConnell}, D.~J., \& {Philip}, A.~G.~D. 1978, \pasp, 90,
  621

\bibitem[{{Sasaki} {et~al.}(2000{\natexlab{a}}){Sasaki}, {Haberl}, \&
  {Pietsch}}]{2000A&AS..143..391S}
{Sasaki}, M., {Haberl}, F., \& {Pietsch}, W. 2000{\natexlab{a}}, \aaps, 143,
  391

\bibitem[{{Sasaki} {et~al.}(2000{\natexlab{b}}){Sasaki}, {Haberl}, \&
  {Pietsch}}]{2000A&AS..147...75S}
{Sasaki}, M., {Haberl}, F., \& {Pietsch}, W. 2000{\natexlab{b}}, \aaps, 147, 75

\bibitem[{{Sasaki} {et~al.}(2002){Sasaki}, {Haberl}, \&
  {Pietsch}}]{2002A&A...392..103S}
{Sasaki}, M., {Haberl}, F., \& {Pietsch}, W. 2002, \aap, 392, 103

\bibitem[{{Skrutskie} {et~al.}(2006){Skrutskie}, {Cutri}, {Stiening},
  {Weinberg}, {Schneider}, {Carpenter}, {Beichman}, {Capps}, {Chester},
  {Elias}, {Huchra}, {Liebert}, {Lonsdale}, {Monet}, {Price}, {Seitzer},
  {Jarrett}, {Kirkpatrick}, {Gizis}, {Howard}, {Evans}, {Fowler}, {Fullmer},
  {Hurt}, {Light}, {Kopan}, {Marsh}, {McCallon}, {Tam}, {Van Dyk}, \&
  {Wheelock}}]{2006AJ....131.1163S}
{Skrutskie}, M.~F., {Cutri}, R.~M., {Stiening}, R., {et~al.} 2006, \aj, 131,
  1163

\bibitem[{{Stanimirovic} {et~al.}(1999){Stanimirovic}, {Staveley-Smith},
  {Dickey}, {Sault}, \& {Snowden}}]{1999MNRAS.302..417S}
{Stanimirovic}, S., {Staveley-Smith}, L., {Dickey}, J.~M., {Sault}, R.~J., \&
  {Snowden}, S.~L. 1999, \mnras, 302, 417

\bibitem[{{Stiele} {et~al.}(2011){Stiele}, {Pietsch}, {Haberl},
  {Hatzidimitriou}, {Barnard}, {Williams}, {Kong}, \&
  {Kolb}}]{2011A&A...534A..55S}
{Stiele}, H., {Pietsch}, W., {Haberl}, F., {et~al.} 2011, \aap, 534, A55

\bibitem[{{Str{\"u}der} {et~al.}(2001){Str{\"u}der}, {Briel}, {Dennerl},
  {Hartmann}, {Kendziorra}, {Meidinger}, {Pfeffermann}, {Reppin}, {Aschenbach},
  {Bornemann}, {Br{\"a}uninger}, {Burkert}, {Elender}, {Freyberg}, {Haberl},
  {Hartner}, {Heuschmann}, {Hippmann}, {Kastelic}, {Kemmer}, {Kettenring},
  {Kink}, {Krause}, {M{\"u}ller}, {Oppitz}, {Pietsch}, {Popp}, {Predehl},
  {Read}, {Stephan}, {St{\"o}tter}, {Tr{\"u}mper}, {Holl}, {Kemmer}, {Soltau},
  {St{\"o}tter}, {Weber}, {Weichert}, {von Zanthier}, {Carathanassis}, {Lutz},
  {Richter}, {Solc}, {B{\"o}ttcher}, {Kuster}, {Staubert}, {Abbey}, {Holland},
  {Turner}, {Balasini}, {Bignami}, {La Palombara}, {Villa}, {Buttler},
  {Gianini}, {Lain{\'e}}, {Lumb}, \& {Dhez}}]{2001A&A...365L..18S}
{Str{\"u}der}, L., {Briel}, U., {Dennerl}, K., {et~al.} 2001, \aap, 365, L18

\bibitem[{{Sturm} {et~al.}(2012{\natexlab{a}}){Sturm}, {Dra{\v{s}}kovi{\'c}},
  {Filipovic}, {Haberl}, {Pietsch}, {Wong}, {de Horta}, {Crawford}, \&
  {Ehle}}]{Sturm_agn}
{Sturm}, R., {Dra{\v{s}}kovi{\'c}}, D., {Filipovic}, M., {et~al.}
  2012{\natexlab{a}}, \aap, to be submitted

\bibitem[{{Sturm} {et~al.}(2011{\natexlab{a}}){Sturm}, {Haberl}, {Coe},
  {Bartlett}, {Buckley}, {Corbet}, {Ehle}, {Filipovi{\'c}}, {Hatzidimitriou},
  {Mereghetti}, {La Palombara}, {Pietsch}, {Tiengo}, {Townsend}, \&
  {Udalski}}]{2011A&A...527A.131S}
{Sturm}, R., {Haberl}, F., {Coe}, M.~J., {et~al.} 2011{\natexlab{a}}, \aap,
  527, A131

\bibitem[{{Sturm} {et~al.}(2011{\natexlab{b}}){Sturm}, {Haberl}, {Greiner},
  {Pietsch}, {La Palombara}, {Ehle}, {Gilfanov}, {Udalski}, {Mereghetti}, \&
  {Filipovi{\'c}}}]{2011A&A...529A.152S}
{Sturm}, R., {Haberl}, F., {Greiner}, J., {et~al.} 2011{\natexlab{b}}, \aap,
  529, A152

\bibitem[{{Sturm} {et~al.}(2012{\natexlab{b}}){Sturm}, {Haberl}, {Pietsch},
  {Ballet}, {Bomans}, {Buckley}, {Coe}, {Corbet}, {Ehle}, {Filipovic},
  {Gilfanov}, {Hatzidimitriou}, {La Palombara}, {Mereghetti}, {Snowden}, \&
  {Tiengo}}]{Sturm_cat}
{Sturm}, R., {Haberl}, F., {Pietsch}, W., {et~al.} 2012{\natexlab{b}}, \aap,
  submitted (SHP12)

\bibitem[{{Sturm} {et~al.}(2012{\natexlab{c}}){Sturm}, {Haberl}, {Pietsch},
  {Coe}, {Mereghetti}, {La Palombara}, {Owen}, \&
  {Udalski}}]{2012A&A...537A..76S}
{Sturm}, R., {Haberl}, F., {Pietsch}, W., {et~al.} 2012{\natexlab{c}}, \aap,
  537, A76

\bibitem[{{T{\"u}llmann} {et~al.}(2011){T{\"u}llmann}, {Gaetz}, {Plucinsky},
  {Kuntz}, {Williams}, {Pietsch}, {Haberl}, {Long}, {Blair}, {Sasaki},
  {Winkler}, {Challis}, {Pannuti}, {Edgar}, {Helfand}, {Hughes}, {Kirshner},
  {Mazeh}, \& {Shporer}}]{2011ApJS..193...31T}
{T{\"u}llmann}, R., {Gaetz}, T.~J., {Plucinsky}, P.~P., {et~al.} 2011, \apjs,
  193, 31

\bibitem[{{T{\"u}llmann} {et~al.}(2009){T{\"u}llmann}, {Long}, {Pannuti},
  {Winkler}, {Plucinsky}, {Gaetz}, {Williams}, {Kuntz}, {Pietsch}, {Blair},
  {Haberl}, \& {Smith}}]{2009ApJ...707.1361T}
{T{\"u}llmann}, R., {Long}, K.~S., {Pannuti}, T.~G., {et~al.} 2009, \apj, 707,
  1361

\bibitem[{{Turner} {et~al.}(2001){Turner}, {Abbey}, {Arnaud}, {Balasini},
  {Barbera}, {Belsole}, {Bennie}, {Bernard}, {Bignami}, {Boer}, {Briel},
  {Butler}, {Cara}, {Chabaud}, {Cole}, {Collura}, {Conte}, {Cros}, {Denby},
  {Dhez}, {Di Coco}, {Dowson}, {Ferrando}, {Ghizzardi}, {Gianotti}, {Goodall},
  {Gretton}, {Griffiths}, {Hainaut}, {Hochedez}, {Holland}, {Jourdain},
  {Kendziorra}, {Lagostina}, {Laine}, {La Palombara}, {Lortholary}, {Lumb},
  {Marty}, {Molendi}, {Pigot}, {Poindron}, {Pounds}, {Reeves}, {Reppin},
  {Rothenflug}, {Salvetat}, {Sauvageot}, {Schmitt}, {Sembay}, {Short},
  {Spragg}, {Stephen}, {Str{\"u}der}, {Tiengo}, {Trifoglio}, {Tr{\"u}mper},
  {Vercellone}, {Vigroux}, {Villa}, {Ward}, {Whitehead}, \&
  {Zonca}}]{2001A&A...365L..27T}
{Turner}, M. J.~L., {Abbey}, A., {Arnaud}, M., {et~al.} 2001, \aap, 365, L27

\bibitem[{{van der Heyden} {et~al.}(2004){van der Heyden}, {Bleeker}, \&
  {Kaastra}}]{2004A&A...421.1031V}
{van der Heyden}, K.~J., {Bleeker}, J.~A.~M., \& {Kaastra}, J.~S. 2004, \aap,
  421, 1031

\bibitem[{{Wang} {et~al.}(1991){Wang}, {Hamilton}, {Helfand}, \&
  {Wu}}]{1991ApJ...374..475W}
{Wang}, Q., {Hamilton}, T., {Helfand}, D.~J., \& {Wu}, X. 1991, \apj, 374, 475

\bibitem[{{Wang} \& {Wu}(1992)}]{1992ApJS...78..391W}
{Wang}, Q. \& {Wu}, X. 1992, \apjs, 78, 391

\bibitem[{{Wilms} {et~al.}(2000){Wilms}, {Allen}, \&
  {McCray}}]{2000ApJ...542..914W}
{Wilms}, J., {Allen}, A., \& {McCray}, R. 2000, \apj, 542, 914

\bibitem[{{Xu} {et~al.}(2005){Xu}, {Zhang}, \& {Han}}]{2005ChJAA...5..165X}
{Xu}, J.-W., {Zhang}, X.-Z., \& {Han}, J.-L. 2005, \cjaa, 5, 165

\bibitem[{{Yokogawa} {et~al.}(2003){Yokogawa}, {Imanishi}, {Tsujimoto},
  {Koyama}, \& {Nishiuchi}}]{2003PASJ...55..161Y}
{Yokogawa}, J., {Imanishi}, K., {Tsujimoto}, M., {Koyama}, K., \& {Nishiuchi},
  M. 2003, \pasj, 55, 161

\bibitem[{{Yokogawa} {et~al.}(2000){Yokogawa}, {Torii}, {Imanishi}, \&
  {Koyama}}]{2000PASJ...52L..37Y}
{Yokogawa}, J., {Torii}, K., {Imanishi}, K., \& {Koyama}, K. 2000, \pasj, 52,
  L37

\bibitem[{{Zaritsky} {et~al.}(2002){Zaritsky}, {Harris}, {Thompson}, {Grebel},
  \& {Massey}}]{2002AJ....123..855Z}
{Zaritsky}, D., {Harris}, J., {Thompson}, I.~B., {Grebel}, E.~K., \& {Massey},
  P. 2002, \aj, 123, 855

\end{thebibliography}

\Online

\addtocounter{table}{-3} 
\begin{table}
\caption{\xmm\ observations of the large programme SMC survey.}
\begin{center}
\begin{tabular}{lllll}
\hline\hline\noalign{\smallskip}
\multicolumn{1}{c}{ID}&
\multicolumn{1}{c}{ObsID}&
\multicolumn{1}{c}{R.A. (J2000)}&
\multicolumn{1}{c}{Dec. (J2000)}&
\multicolumn{1}{c}{Date}\\

\noalign{\smallskip}\hline\noalign{\smallskip}
S1  & 0601210101 & 00:58:16.4 & -71:28:45 & 2009-5-14\\
S2  & 0601210201 & 00:53:58.9 & -71:43:48 & 2009-9-25\\
S3  & 0601210301 & 00:49:01.8 & -71:56:42 & 2009-5-18\\
S4  & 0601210401 & 01:02:59.3 & -71:36:41 & 2009-9-25\\
S5  & 0601210501 & 00:58:54.9 & -71:49:49 & 2009-9-25\\
S6  & 0601210601 & 00:55:20.2 & -72:01:42 & 2009-9-27\\
S7  & 0601210701 & 00:49:28.7 & -72:17:15 & 2009-9-27\\
S8  & 0601210801 & 00:56:15.5 & -72:21:55 & 2009-10-9\\
S9  & 0601210901 & 00:48:08.9 & -72:37:39 & 2009-9-27\\
S10 & 0601211001 & 00:43:56.7 & -72:44:38 & 2009-11-9\\
S11 & 0601211101 & 00:42:29.5 & -73:02:27 & 2009-10-18\\
S12 & 0601211201 & 00:42:25.2 & -73:20:11 & 2009-10-20\\
S13 & 0601211301 & 00:46:28.7 & -73:24:25 & 2009-10-3\\
S14 & 0601211401 & 00:52:19.2 & -73:09:03 & 2009-11-4\\
S15 & 0601211501 & 00:56:25.5 & -73:02:58 & 2009-10-13\\
S16 & 0601211601 & 00:58:21.5 & -72:48:27 & 2009-10-11\\
S17 & 0601211701 & 01:02:23.1 & -72:35:23 & 2009-10-16\\
S18 & 0601211801 & 01:04:04.7 & -72:22:53 & 2009-11-13\\
S19 & 0601211901 & 01:08:33.2 & -72:09:54 & 2009-11-30\\
S20 & 0601212001 & 01:12:56.4 & -72:22:38 & 2009-11-27\\
S21 & 0601212101 & 01:11:32.3 & -72:43:31 & 2009-11-16\\
S22 & 0601212201 & 01:13:35.3 & -73:01:05 & 2009-11-19\\
S23 & 0601212301 & 01:17:03.4 & -73:04:05 & 2009-9-9\\
S24 & 0601212401 & 01:20:47.6 & -73:15:35 & 2009-6-29\\
S25 & 0601212501 & 01:12:31.5 & -73:18:24 & 2009-9-9\\
S26 & 0601212601 & 01:08:33.1 & -72:54:46 & 2009-6-29\\
S27 & 0601212701 & 01:07:54.8 & -73:09:25 & 2009-12-26\\
S28 & 0601212801 & 01:01:54.0 & -73:07:05 & 2009-12-7\\
S29 & 0601212901 & 00:57:04.8 & -73:20:23 & 2009-9-13\\
S30 & 0601213001 & 00:53:18.3 & -73:32:45 & 2009-9-13\\
S31 & 0601213201 & 01:02:23.1 & -72:35:23 & 2010-3-12\\
S32 & 0601213301 & 01:01:54.0 & -73:07:05 & 2010-3-12\\
S33 & 0601213401 & 00:58:16.4 & -71:28:45 & 2010-3-16\\

\noalign{\smallskip}\hline
\end{tabular}
\end{center}
\label{tab-obs}
\end{table}

\addtocounter{table}{-1} 
\begin{table}
\caption{Continued: archival and calibration (1E0102.2-7219) observations.}
\begin{center}
\begin{tabular}{lllll}
\hline\hline\noalign{\smallskip}
\multicolumn{1}{c}{ID}&
\multicolumn{1}{c}{ObsID}&
\multicolumn{1}{c}{R.A. (J2000)}&
\multicolumn{1}{c}{Dec. (J2000)}&
\multicolumn{1}{c}{Date}\\

\noalign{\smallskip}\hline\noalign{\smallskip}
A1  & 0112780201 & 00:59:13.0 & -71:38:50 & 2000-9-19\\
A2  & 0110000101 & 00:49:07.0 & -73:14:06 & 2000-10-15\\
A3  & 0110000201 & 00:59:26.0 & -72:10:11 & 2000-10-17\\
A4  & 0110000301 & 01:04:52.0 & -72:23:10 & 2000-10-17\\
A6  & 0011450101 & 01:17:05.1 & -73:26:35 & 2001-5-31\\
A9  & 0018540101 & 00:59:26.8 & -72:09:55 & 2001-11-20\\
A10 & 0084200101 & 00:56:41.7 & -72:20:24 & 2002-3-30\\
A11 & 0142660801 & 00:59:26.4 & -71:18:48 & 2003-11-17\\
A12 & 0157960201 & 00:55:22.0 & -72:42:00 & 2003-12-18\\
A14 & 0212282601 & 00:59:26.8 & -72:09:54 & 2005-3-27\\
A15 & 0304250401 & 00:59:26.8 & -72:09:54 & 2005-11-27\\
A16 & 0304250501 & 00:59:26.8 & -72:09:54 & 2005-11-29\\
A17 & 0304250601 & 00:59:26.8 & -72:09:54 & 2005-12-11\\
A20 & 0301170101 & 01:08:06.4 & -72:52:23 & 2006-3-22\\
A21 & 0301170201 & 00:52:12.1 & -72:01:42 & 2006-3-23\\
A22 & 0301170601 & 00:40:23.8 & -72:46:50 & 2006-3-27\\
A23 & 0301170301 & 00:42:46.6 & -73:35:38 & 2006-4-6\\
A24 & 0402000101 & 01:03:52.2 & -72:54:28 & 2006-10-3\\
A25 & 0404680101 & 00:47:36.0 & -73:08:24 & 2006-10-5\\
A26 & 0404680201 & 00:52:26.4 & -72:52:12 & 2006-11-1\\
A27 & 0403970301 & 00:47:39.4 & -72:59:31 & 2007-3-12\\
A28 & 0404680301 & 00:51:00.7 & -73:24:17 & 2007-4-11\\
A29 & 0404680501 & 01:07:42.3 & -72:30:11 & 2007-4-12\\
A32 & 0500980101 & 00:53:02.4 & -72:26:17 & 2007-6-23\\
A33 & 0503000201 & 00:48:23.4 & -73:41:00 & 2007-10-28\\
A34 & 0503000301 & 00:40:23.8 & -72:46:50 & 2008-3-16\\
A35 & 0656780101 & 00:49:05.9 & -72:50:55 & 2010-3-24\\
A36 & 0656780201 & 00:49:05.9 & -72:50:55 & 2010-3-27\\
A37 & 0656780301 & 00:49:05.9 & -72:50:55 & 2010-3-30\\
C1  & 0123110201 & 01:03:50.0 & -72:01:55 & 2000-4-16\\
C2  & 0123110301 & 01:03:50.0 & -72:01:55 & 2000-4-17\\
C3  & 0135720601 & 01:03:50.0 & -72:01:55 & 2001-4-14\\
C4  & 0135720801 & 01:04:00.0 & -72:00:16 & 2001-12-25\\
C5  & 0135720901 & 01:04:01.7 & -72:01:51 & 2002-4-20\\
C8  & 0135721301 & 01:03:56.4 & -72:00:28 & 2002-12-14\\
C9  & 0135721401 & 01:04:18.1 & -72:02:32 & 2003-4-20\\
C10 & 0135721501 & 01:03:45.6 & -72:01:07 & 2003-10-27\\
C11 & 0135721701 & 01:03:45.6 & -72:01:07 & 2003-11-16\\
C12 & 0135721901 & 01:04:17.3 & -72:02:39 & 2004-4-28\\
C13 & 0135722401 & 01:03:45.6 & -72:01:07 & 2004-10-14\\
C14 & 0135722001 & 01:04:03.6 & -72:01:44 & 2004-10-26\\
C15 & 0135722101 & 01:03:59.6 & -72:01:44 & 2004-11-7\\
C18 & 0135722501 & 01:04:17.3 & -72:02:39 & 2005-4-17\\
C19 & 0135722601 & 01:03:47.1 & -72:00:57 & 2005-11-5\\
C20 & 0135722701 & 01:04:01.7 & -72:01:51 & 2006-4-20\\
C21 & 0412980101 & 01:03:47.1 & -72:00:57 & 2006-11-5\\
C22 & 0412980201 & 01:04:01.7 & -72:01:51 & 2007-4-25\\
C23 & 0412980301 & 01:03:47.1 & -72:00:57 & 2007-10-26\\
C24 & 0412980501 & 01:04:01.7 & -72:01:51 & 2008-4-19\\
C25 & 0412980701 & 01:04:01.7 & -72:01:51 & 2008-11-14\\
C26 & 0412980801 & 01:04:01.7 & -72:01:51 & 2009-4-13\\
C27 & 0412980901 & 01:04:01.7 & -72:01:51 & 2009-10-21\\
C28 & 0412981001 & 01:04:01.7 & -72:01:51 & 2010-4-21\\

\noalign{\smallskip}\hline
\end{tabular}
\end{center}
\end{table}

\end{document}